# Controlling Competing Feedback Mechanisms for Programmable Patterns via Nonlinear Laser Lithography

Özgün Yavuz[1,†], Ihor Pavlov[2], Abdullah bin Aamir[1], Sezin Galioglu[3], Petro Deminskyi[4], Ü. Seleme Nizam[1], Onur Tokel[3,4], Serim Ilday[1,5], and F. Ömer Ilday[1,5,†]

[1] Faculty of Electrical Engineering and Information Technology, Ruhr-Universität Bochum, 44801 Bochum, Germany

[2] Department of Physics, Middle East Technical University, 06800 Ankara, Turkey

[3] UNAM-National Nanotechnology Research Center, Bilkent University, 06800 Ankara, Turkey

[4] Department of Physics, Bilkent University, 06800 Ankara, Turkey

[5] Faculty of Physics and Astronomy, Ruhr-Universität Bochum, 44801 Bochum, Germany

† Corresponding author: oezguen.yavuz@ruhr-uni-bochum.de & oemer.ilday@ruhr-uni-bochum.de

Controlling laser-induced pattern formation remains a long-standing challenge. A key advance was recognising the pivotal role of intrinsic feedback mechanisms in self-organisation, which enabled self-similar patterns with long-range order through nonlinear laser lithography. This concept was recently leveraged to surpass the diffraction limit. However, demonstrated structures in laser-induced pattern formation have been relatively simple and material-specific, as they rely on a single assumed formation mechanism for each material. Here, we reveal the coexistence of competing feedback mechanisms that drive distinct chemical and ablative processes, resulting in patterns with different symmetries. We show that one mechanism can be selectively activated, while completely suppressing the other, by tuning the laser parameters independently of the material. The competing mechanisms break surface symmetries in distinct lateral and vertical directions, and the selection can be dynamically inverted *in situ*, leading to rich surface chemistries and morphologies. We demonstrate the potential for applications through the maskless and ambient-air fabrication of complex composite patterns on diverse materials and substrates. These include seamlessly stitched superhydrophilic and superhydrophobic domains, structural colour, and a holographic temperature sensor. Combining programmable mechanism selection with recent demonstrations of sub-10 nm features could enable the maskless fabrication of composite structures that rival the capabilities of nanofabrication techniques.

Pattern formation is historically the first[1] and the most studied[2-5] self-organised process driven by a laser. However, the resulting structures suffered from poor long-range order until the advent of nonlinear laser lithography (NLL) in 2013[6], which introduced deliberately exploiting intrinsic feedback mechanisms between the laser beam and the material's changes, based on an analogy to how self-amplitude modulation drives mode-locking of lasers. This conceptual advance has led to patterns with sub-nanometer Allan deviation[6], and opened the door to 3D geometries[3,7,8] and structures that surpass the diffraction limit[4,9]. Here, we present another conceptual advance where multiple competing pattern formation mechanisms are shown to coexist, with one mechanism being selectively chosen at the complete exclusion of the other, enabling programmable patterns with rich surface chemistries and morphologies that could be used for various technological purposes. We provide proof of concept demonstrations for a holographic sensor, wettability control, and structural colouring of large areas and flexible surfaces.

One mechanism forms patterns by ablation of the surface[11-14], which occurs only at points where the local intensity exceeds the ablation threshold due to the interference between the incident laser beam and the surface plasmon-polariton (SPP) waves excited by the beam. This results in stripe-like structures that are depressions below the surface, oriented perpendicular to the laser's polarisation[6,10]. For the second mechanism, the structures look similar but grow outward from the surface by a thermochemical reaction at the constructive interference points[6]. In ambient atmosphere, this results in oxidation; however, fundamentally, the reaction could be any kinetically favourable process occurring in a suitable environment. The threshold intensity depends strongly on the cumulative number of incident pulses because the process is thermally activated. The resulting stripes are parallel[6,9] to the laser polarisation. Interestingly, ablation- and oxidation-type (thermochemical) patterns have never been observed on the same surface but were regarded as unrelated phenomena that do not co-exist. This has led to speculation on the conditions where SPPs or retarded potentials are independently responsible.

We start by demonstrating that these two pattern formation mechanisms are mutually *orthogonal* and *tightly feedback-driven*, ensuring *nonlocal*, self-similar pattern formation extending beyond the local interaction volume. Then, we show that the pattern formation process can be controlled through *intrinsic* feedback mechanisms, which enables various *device applications*. We elaborate on each of these points by showing that NLL supports both mechanisms simultaneously, which compete, with the marginally stronger one completely excluding the other due to self-strengthening effect of the intrinsic feedback. We then show that the selected mechanism can be switched, inverting the symmetries of the resulting pattern and altering their morphological and chemical compositions simply by adjusting the laser parameters. Using large numbers of low-energy, high-repetition-rate pulses, we form oxidation-based structures, NLL $_{\parallel}$, that grow out of the surface and are oriented parallel to the laser polarisation (Figure 1a). In contrast, a small number of high-energy pulses create ablation-based structures, $NLL_{\perp}$, that grow into the surface and are oriented perpendicular to the polarisation (Figure 1b).

$NLL_{\parallel}$ and $NLL_{\perp}$ break the symmetries in two distinct ways: first, the stripe-like structures they form are mutually orthogonal, and second, they grow either out of or into the surface, thus breaking the symmetry transverse to the surface. We demonstrate both types on various materials, from metals to a semiconductor (Figure 2). Finally, we highlight potential technological applications (Figure 3) of composite patterns with varying morphologies achieved through programmable switching of the $NLL_{\parallel}$ and $NLL_{\perp}$ patterns. We conclude by suggesting that competing mechanisms are likely behind recent demonstrations of laser-driven structures and that programmable mechanism selection introduced here could be applicable to them.

**The pattern forming mechanisms**

The patterns are formed by focussing an ultrafast laser beam on a surface, where every illuminated point infinitesimally scatters the laser light in proportion to its local height differences[6] (Figure 1c). Part of the scattered light propagates along the surface. As long as the focal spot is smaller than the propagation distance of the scattered light, all scatterers interfere fully coherently, thus collectively determining the total light intensity at a given point, which renders the feedback strongly nonlocal[6]. The total intensity along the surface, $I_{\mathrm{total},n}$, is described by a 2D convolution integral (Figure 1d),

$$I_{\mathrm{total},n}(x,y) = \left| E_n(x,y) + \zeta \iint h_n(x',y') E_n(x',y') K(x-x', y-y') \mathrm{d}x' \mathrm{d}y' \right|^2, \tag{1}$$

where $E_n$ is the electric field due to the $n^{\mathrm{th}}$ pulse impinging on the material surface, $h_n$ is the surface height profile, $K$ is the scattering kernel, and $\zeta$ is the scattering strength. The scattering occurs via two orthogonal processes: radiation remnants, including dipole radiation (creating $\mathrm{NLL}_{\parallel}$), and SPP excitation (creating $\mathrm{NLL}_{\perp}$). Both are included in the scattering kernel (see Methods for the full model description).

Numerical simulations account for every surface point, but here we simplify the description by assuming that the first laser pulse scatters from a single point (Figure 1c(i)), modifying the surface according to one of the mechanisms. The light intensity scattered from a single defect is plotted in Figures 1c(i) and 1c(ii) to visualise the kernel. The two scattering processes have different amplitudes and decay lengths, as seen from the cross-sections of the intensity distribution parallel (green curve) and perpendicular (blue curve) to the polarisation (Figure 1c(iii)). An incident laser beam polarised along the $y$-axis has a long-range intensity modulation along the $x$-axis mainly due to dipole radiation, resulting in structures parallel to the polarisation ($\mathrm{NLL}_{\parallel}$). On the other hand, the component along the $y$-axis is primarily due to SPP coupling, which has higher peak amplitudes but with a shorter-range intensity modulation, leading to structures perpendicular to to the polarisation ($\mathrm{NLL}_{\perp}$). These minor differences, amplified exponentially by the nonlocal feedback, form the selection mechanism, as explained below. However, before moving further, it is important to highlight a key point. There is no fundamental reason why the SPP field must exclusively drive ablation (*i.e.*, $\mathrm{NLL}_{\perp}$) while dipole radiation drives the thermochemical reaction (*i.e.*, $\mathrm{NLL}_{\parallel}$). This distinction arises from a practical consequence: for all the materials examined in this study, the SPP field produces higher but narrower intensity distribution that surpass the ablation threshold before reaching the threshold for the chemical reaction. Conversely, the broader intensity distribution associated with dipole radiation crosses the chemical reaction threshold before exceeding the ablation threshold (Figure S1). If the thresholds were reversed, the SPP field would drive $\mathrm{NLL}_{\parallel}$, and dipole radiation would drive $\mathrm{NLL}_{\perp}$.

**The origin of nonlocal feedback and competition between the mechanisms**

The unprocessed surface is typically not pristine nor atomically flat, instead, there are numerous random height variations, each acting as a scatterer, initially forming a random interference profile. At points of constructive interference, the surface will remain unchanged or be modified by oxidation (blue path in Figure 1d) or ablation (green path in Figure 1d), depending on which, if either, threshold is exceeded. The step-by-step formation of $NLL_{\parallel}$ and $NLL_{\perp}$ is illustrated in Figure 1e via numerical simulations. Since the power of the laser beam is highest at its centre, modification begins and spreads from there (Figure 1e). Depending on the local intensity and the number of pulses per spot, the ablation or oxidation threshold may be reached, leading to the formation of $NLL_{\perp}$ or $NLL_{\parallel}$, respectively. The upper row of Figure 1e corresponds to $NLL_{\parallel}$, which is formed gradually by thousands of high-repetition-rate, low-energy pulses. In contrast, $NLL_{\perp}$ forms more abruptly through a small number of high-energy pulses. In either case, the local height changes by $\Delta h_n(x, y)$, increasing for oxidation and decreasing for ablation, and the surface profile is updated to $h_{n+1}(x, y)$. When the next pulse arrives, its intensity is scattered differently by the updated height profile, completing the feedback loop.

Importantly, either mechanism could only marginally outcompete the other without the nonlocal feedback. Ordinarily, the weaker mechanism would also modify the surface, albeit to a lesser extent, leading to inhomogeneous and distorted patterns. This is where the critical role of nonlocal feedback, arising from the convolution integration in equation 1, comes into play to achieve complete competitive exclusion: Once the smallest protostructures of the dominant mechanism form (at $t_2$ and onwards, Figure 1e), they exponentially strengthen the chosen mechanism by altering $I_{\mathrm{total},n}$, which sets the rate of change, $\Delta h_n$. This asymmetric growth results in pristine patterns (Figure S2) that show no sign of the other mechanism, which is effectively hidden (Video 1 and 2). Scanning electron microscope (SEM) images of fully formed $NLL_{\parallel}$ and $NLL_{\perp}$ patterns on the same Ti sample are shown in Figure 1e.

Given the exponential reinforcement by nonlocal feedback, one might expect that switching the mechanism, once chosen, is impossible. However, it can easily be achieved by momentarily turning off the pulses, moving the focal spot beyond the existing structures so they no longer provide nonlocal feedback, and then turning the pulses back on with the new parameters. This process inverts the symmetries, and any gaps between the two types of structures can be filled by scanning the beam over the surface. As discussed in the further sections, the symmetries can be inverted as many times as desired, which helped demonstrate the device applications presented in Figure 3.

**The role of intrinsic feedback mechanisms**

Two mechanisms are not equivalent; there are two drawbacks to the $NLL_{\perp}$ process compared to $NLL_{\parallel}$. The first is the relatively abrupt formation process via a small number of laser pulses. This results in large height changes per pulse, so the feedback between the evolving surface profile and the redistribution of the light intensity for each pulse occurs in fewer and larger steps. This can be avoided by ablation via bursts containing thousands of pulses (data not shown here). In the case of oxidation, in addition to nonlocal positive feedback, negative feedback causes saturation of the structure's growth after it reaches a certain height (Figures 1d and S2). This feedback is based on the self-limitation of oxygen access to the growing structures proportional to their height. Practically, it makes the process immune to multiple scans and removes the lower limit for the scanning speed.

Importantly, $NLL_{\parallel}$ exhibits intrinsic negative feedback (blue path in Figure 1d). As the oxide grows thicker, it gets progressively harder for $O_2$ from ambient air to reach the activated material beneath, thus limiting further growth. The self-limiting nature of the process results in a well-defined and uniform height[6] (Figure S2). The SEM images of oxidation-based parallel surface structures formed on the Ti surface are presented in Figure S3a, where the laser beam was scanned over the same area multiple times. There is no recognisable difference between structures in terms of their quality. In contrast, when the formation is governed by ablation, it does not have inherent negative feedback to halt the structure formation, which is the second drawback of $NLL_{\perp}$. Each pulse impinging on the surface ablates more material and destroys the surface structures. The absence of negative feedback makes the $NLL_{\perp}$ process prone to disruption by multiple scans. Even after a second scan, the quality of the structure decreases significantly and is ruined after the fifth scan (Figure S3b). This requires requires additional precautions to preserve the quality of the structure and makes it more challenging to obtain highly uniform structures via $NLL_{\perp}$. When the surface is a thin film on a substrate, ablation self-terminates when the thin film is fully removed, but for thicker films or bulk targets, limiting the number of pulses is essential (Figure S3). For example, the optimal number of pulses per spot is optimised to be 4-10 for the Ti sample.

**Universality of the mechanism selection and symmetry inversions**

The mechanism of competition, selection, and controllable symmetry inversions in $NLL_{\parallel}$ and $NLL_{\perp}$ is universal in that the dynamical steps discussed above and qualitative features are not material specific, provided the underlying physical processes are supported in that material. Light scattering, which drives both $NLL_{\parallel}$ and $NLL_{\perp}$, occurs via laser-induced (dipole) radiation and SPP. The former applies universally to any solid surface, while the latter is supported by most non-dielectrics. Similarly, ablation is possible in all materials, and ultrafast oxidation occurs in many, though not all. These points are further discussed below. As mentioned earlier, while we focus on oxidation due to its reliance only on ambient air, the process can involve other chemical reactions, such as nitrate or carbide formation, if relevant active gases are introduced during the formation process.

To demonstrate this universality, we created structures on various metals such as Ti, Co, Al, Ni, Nb, Cr, and a semiconductor, Si, despite their widely different absorptions, refractive indices, and heat diffusion rates. $NLL_{\parallel}$ patterns on all these materials are shown in Figure 2a, in addition to flexible glass and polymer Kapton tape featured in Figure 3. The strong and narrow reciprocal peaks shown in the lower row of Figure 2a indicate high uniformity. $NLL_{\perp}$ patterns on the same sample and their reciprocal peaks are also presented. Notably, $NLL_{\parallel}$ structures exhibit superior quality and periodicity, especially over large areas, due to their more gradual growth and self-limiting negative feedback. However, the abruptness of $NLL_{\perp}$ is not inevitable. Future work may achieve more gradual ablation by employing bursts containing thousands of pulses[16].

Next, we demonstrate that the chosen mechanism can be deliberately switched to the other over the same material surface. This transition is seamless, as shown in Figure 2b (Figure S4 and Videos 3-5). We used $6\times10^4$ to $6\times10^5$ pulses per spot to form $NLL_{\parallel}$ with pulse fluences ranging from 30 to 45 mJ/cm$^2$, depending on the material. In contrast, $NLL_{\perp}$ required only 6 to 60 pulses at the higher fluences of 200 to 290 mJ/cm$^2$ (see Table 1 in the Methods section). The processed samples are diced and their cross-section profiles are imaged using SEM to confirm that $NLL_{\parallel}$ structures grow out of the surface (Figures 2b(i) and 2b(iii)) while $NLL_{\perp}$ structures grow into the surface via ablation (Figures 2b(ii) and 2b(iv)). The surface heights are shown with red dashed lines, which denote the unprocessed parts on the surface (outside the field of view). These measurements verify that $NLL_{\parallel}$ structures grow above the surface, while $NLL_{\perp}$ structures are depressions below it (see also Figure S4).

Our theory predicts both the range of pulse parameters (Figure 2c) and the periodicities of the resulting structures (Figures 2d, S5 and S6). For $NLL_{\perp}$, the period corresponds to that of the SPP waves[13]. Predicting the $NLL_{\parallel}$ periods is more challenging, but calculations assuming an effective index $n_{\text{eff}}$, to account for the periodic metal and metal-oxide composition of the surface[17] show good agreement with the experiments. The uncertainties in the predicted values stems from limited knowledge or variability in material properties, which significantly influence our model's sensitivity.

**Proof-of-concept device applications**

The universality of the formation mechanisms, combined with the ability to switch between them at will, offers far greater possibilities than relying on a single mechanism. These possibilities extend beyond controlling the symmetries; they include modifying the morphology with (*i*) structures growing into or out of the surface, (*ii*) their periodicity, (*iii*) their chemistry (the native material or its oxide), and (*iv*) by tiling them over the surfaces in differing orientations and periodicities. Armed with the predictive power of our theoretical model, we showcase several technological applications. First, we present a functional holographic temperature sensor (Figures 3a and 3b, and Video 6). Next, we demonstrate advanced control over surface wettability (Figure 3c, and Videos 7 and 8). Finally, we introduce programmable sequences of repeated symmetry inversions that tile complex surface patterns for structural colouring (Figures 3, S7-S10, and Video 9). These applications were designed to highlight versatility across diverse substrate materials: Kapton tape for the holographic temperature sensor, Si wafers for wettability control, and flexible glass for structural colouring.

The temperature sensor uses a hologram, created over a 100-nm-thick Co-coated Kapton tape (Figure 3a). The sample was placed over a Peltier element to control its temperature electrically and illuminated with white light for detection. As the temperature increased from 27°C to 100°C, then decreased to 37°C, the periodicity of the structures changed due to the thermal expansion and contraction of the tape. This shift modulated the amount of red light detected (Figure 3b and Video 6).

For wettability control, we patterned Si wafers with $NLL_{\perp}$, producing a superhydrophilic surface with a contact angle of 11.1° (Figure 3c(i)). Coating the patterns with a 10-nm-thick $C_4F_8$ polymer rendered the surface superhydrophobic with a contact angle of 179.9° (Figure 3c(ii)). Selective NLL processing then reverted the treated areas to superhydrophilic, now with a contact angle of 19.1°. The fabrication process is both simple, single-step laser scanning, and maskless. Time-lapse images of a water droplet bouncing over the superhydrophilic surface are shown in Figure 3c(iii) (Video 7). Although laser-based wettability control has been reported[18], to our knowledge, this is the first maskless demonstration to achieve transitions between superhydrophobic and superhydrophilic regions over just a few microns on the same surface (Figure 3c(iv)). To showcase the capability, we demonstrate guided water flow along a predetermined path (Video 8).

Structural colouring is also achieved by creating heptomino patterns with four different NLL patterns (Figure 3d). The perceived colours arise entirely from the nanostructure; each tile diffracts a different wavelength range to a specific angle under a white light illumination (see Figures S7-S11). Inspired by the *Limenitis arthemis astyanax* butterfly, we replicated its image by structurally colouring a 4×4 $cm^2$ area on Co-coated flexible glass (Figures 3e and S12). The butterfly's colours were achieved using different laser polarisation states and formation mechanisms. The colours shift with the viewing angle, similar to the natural flapping wings behaviour of the *Limenitis arthemis astyanax* (Video 9).

Future possibilities include expanding the colour palette by varying the angle of polarised light, combining structural colouring with other functionalities, such as wettability and tribological control as demonstrated here and previously[19,20]. Additionally, control of the structural periods can be tuned *in-situ* by changing the incidence angle of the laser beam (Figures S5 and S6) or the laser wavelength. All results reported here were achieved in ambient conditions, where oxygen is the most reactive specie; however, alternative chemical reactions can be exploited under controlled atmospheres, in liquid solutions[21], or using reactants from a plasma jet[22].

**Conclusions and future outlook**

We have demonstrated the control of competing pattern forming mechanisms in NLL with high precision, enabling the selective activation of one mechanism over another through changing the relevant laser parameters. More importantly, the excluded mechanism is suppressed by nonlocal feedback but can be reintroduced by tuning the laser parameters, offering a unique means of manipulating surface patterning by controlling feature size, orientation, or periodicity.

While competitive exclusion ensures high-quality and diverse structures, it complicates the identification of other mechanisms if only a limited range of laser parameters is explored. This new perspective prompted us to revisit recent findings on self-organised structures in semiconductors and transparent dielectrics[3,4,7-9,23,24]. Each demonstration reported distinct mechanisms, often in the same material[3,7,8,23], using widely varying laser parameters, such as nanosecond or femtosecond pulses (resulting in higher[7] or lower[23] refractive indices both in Si) or different wavelengths[7,8]. We conclude that multiple mechanisms likely coexisted but remained concealed due to competitive exclusion.

This new capability represents a pivotal opportunity[25] to combine the distinct mechanisms behind these recent results[4,7,8,23] to enable the maskless, low-cost fabrication of feedback-driven programmable structures with the uniformity, complexity, and potentially sub-10 nm feature sizes[4] to rival top-down nanofabrication techniques.

Beyond technological implications, we have also employed competing feedback mechanisms in NLL as a physical model system to explore a long-standing fundamental question: the origin of homochirality. Competitive exclusion, a well-known phenomenon in ecology[15], describes how a stronger competitor eliminates a weaker one when vying for the same resources—in this case, the opportunity to reshape the surface profile that provides the feedback. In a parallel study[26], we demonstrated that spatially extended systems governed by nonlocal nonlinearities, constituting a form of positive feedback, can exponentially amplify minute differences in growth rates or concentrations, allowing one chiral form to dominate. By extending Frank's classic model of homochirality[27] to include even the simplest nonlocal interactions, we showed that homochirality can emerge under much less restrictive conditions than previously thought, making it a more plausible outcome in prebiotic environments. Together, these results reinforce the dual role of NLL as both a powerful fabrication technique and a versatile physical model system for exploring fundamental principles of self-organisation and symmetry breaking.

## Methods

### Experimental Setup

The setup consists of a home-built, laser-coupled transmission/reflection mode microscope. The schematic drawing of the setup is provided in Figure 2 of Ref. [6]. This setup enables simultaneous diffraction-limited optical imaging during laser-driven pattern formation. The microscope is based on a Nikon Eclipse Ti2 inverted microscope body, equipped with objectives of varying magnifications (Nikon CFI Plan Apo VC 60XC WI, CFI Plan Fluor 40X, CFI S Plan Fluor ELWD 20XC), and a motorised dual-axis translational stage (Standa 8MTF-200 or Thorlabs, MLS203 series). Laser pulses are provided by a Spectra-Physics Spirit One 1040-8-SHG ultrafast laser operating at 1 μm wavelength with a pulse duration of 300 fs. At a repetition rate of 1 MHz, the maximum pulse energy that we used was 1 μJ. The repetition rate of the laser is electronically reducible from 1 MHz down to nearly a single pulse operation.

In nearly all of the experiments, the beam was kept stationary, and the sample was translated via 2D motorised translation stages. Only in experiments requiring faster processing to cover large areas, a 2D galvo scanner (Scanlab, SCANcube 14) was temporarily integrated. A galvo scanner quickly covers areas as large as 5.0×5.0 $cm^2$, with a 600 mm/s scanning speed and the spot size of 12 μm (half-width at $1/e^2$ intensity threshold), corresponding to a processing speed up to 7.2 $mm^2$/s. This is not a hard limit, certainly not of the NLL process. Processing speed can be significantly increased by using, for example, resonant-galvo scanner technology and a setup optimised for high-speed processing.

The illumination was from white or blue LEDs, operating at 450 nm in the latter case. Glass substrates with thin (a few 100 nm) metal films were transparent enough, but bulk samples or Si wafers required the reflection mode. In either imaging mode, the pattern formation can be observed in real time. Videos of superhydrophobicity were captured with an iPhone X camera, while the temperature sensor videos were recorded with a Nikon 5D Mark IV DSLR camera.

### Sample Preparation

A thermal evaporator and RF sputtering device were employed to coat various substrates with different materials. For the co-existence experiments, 150-μm-thick microscope slides were used as the substrate (see Table 1 below for detailed processing parameters and sample thicknesses). To create Figure 2b, we used thin-film coatings of 400-nm Ti and 110-nm Co deposited on a glass slide via thermal sputtering. Then, we patterned parts of the same sample with both $NLL_{\parallel}$ and $NLL_{\perp}$. Lastly, we cut the sample with a precision saw, bisecting the patterns, which were then imaged via scanning electron microscope (FEI Nova NanoSEM).

The temperature sensor was prepared by depositing a 110-nm-thick Co layer on Kapton tape via thermal evaporation. The butterfly pattern was drawn on a 110-nm-thick Co-coated Corning EAGLE XG Slim flexible glass. The thickness of these films was determined using a variable angle spectroscopic ellipsometer (V-VASE, J.A. Woollam Co. Inc., Lincoln, NE) coupled with a rotating analyser and xenon light source. The wettability control was achieved through either a single-step or a three-step fabrication process. The first step is to process the surface with $NNL_{\perp}$. By patterning the surface, we rendered it superhydrophilic with a contact angle of 11.1°. If only a superhydrophilic surface is required, this single step is sufficient because the native wettability characteristics of the unprocessed areas are measured to have a contact angle of ~70° for Si and ~55° for glass. For more complex patterns featuring alternating superhydrophilic and superhydrophilic regions, as in Figure 3c(iv) or Video 8, two additional steps are followed. The second step is to coat the surface with a layer of octafluorocyclobutane ($C_4F_8$) using an inductively coupled plasma system. The third and final step is to convert selected parts of the fully $C_4F_8$-covered and superhydrophobic surface back to superhydrophilic by maskless processing with NLL patterning, resulting in a low contact angle of 19.1°. Although we focussed on Si (100) wafers for their technological significance, the process is entirely applicable to the other materials for which we demonstrated NLL.

**Experimental parameters for the switching experiments**

The parameters for the experiments shown in Figure 2a are given in Table 1. They were performed at a laser repetition rate of 1 MHz and 2 kHz, respectively. In all experiments, the sample was raster-scanned with a scan step of 1 µm, and the focussed spot size was 6 µm ($1/e^2$ width). While the pulse fluence required to form $NLL_{\parallel}$ patterns varies between 35 to 42 mJ/cm², creating $NLL_{\perp}$ patterns requires those values to be between 210 and 350 kJ/m². Similarly, in $NLL_{\perp}$ patterns, the pulses per spot range from 6 to 60, while $NLL_{\parallel}$ we used between $6\times10^4$ to $6\times10^5$ pulses. The only outlier is Si, where fewer pulses were sufficient. This may be related to the laser wavelength being at the edge of the band gap of Si.

**Table 1.** Laser and scanning parameters, and thin-film thicknesses for the performed experiments.

| | $NLL_{\parallel}$ | | | | $NLL_{\perp}$ | | | | |
|---|---|---|---|---|---|---|---|---|---|
| **Material** | **Scanning speed (µm/s)** | **Number of pulses per spot** | **Pulse fluence (mJ/cm²)** | **Peak intensity (TW/cm²)** | **Scanning speed (µm/s)** | **Number of pulses per spot** | **Pulse fluence (kJ/m²)** | **Peak intensity (TW/cm²)** | **Sample thickness (nm)** |
| Cobalt | 30 | 200.000 | 42,4 | 0,159 | 200 | 60 | 212 | 0,797 | 110 |
| Niobium | 20 | 300.000 | 42,4 | 0,159 | 200 | 60 | 212 | 0,797 | 120 |
| Nickel | 10 | 600.000 | 35,3 | 0,133 | 250 | 48 | 283 | 1,064 | 230 |
| Chromium | 30 | 200.000 | 42,4 | 0,159 | 200 | 60 | 212 | 0,797 | 177 |
| Aluminum | 10 | 600.000 | 42,4 | 0,159 | 200 | 60 | 248 | 0,932 | 200 |
| Titanium | 20 | 300.000 | 39,6 | 0,148 | 250 | 12 | 340 | 2,127 | 135 |
| Silicon | 10 | 300 | 707 | 1,326 | 250 | 12 | 707 | 2,216 | 200 |

**Theoretical Model**

The model depicted in Figure 1d begins by describing the scattering of laser light from the surface, followed by the two alternative surface modification mechanisms: ablation and a thermochemical reaction (dominantly oxidation in the ambient atmosphere). When a laser pulse impinges on the sample surface, every infinitesimal point on the surface scatters the electromagnetic wave. The strength of this scattering depends on the local surface properties, particularly whether a given point is elevated or depressed relative to its immediate surroundings. For instance, a point that is merely nanometres taller or lower than its neighbouring points will scatter more effectively than a point that is of equal height to its surroundings. We refer to such points simply as "defects."

The total electric field, $\mathbf{E}_{\mathrm{tot}}(\mathbf{r})$, on the surface can be written as the sum of the incident electric field, $\mathbf{E}_{\mathrm{inc}}(\mathbf{r})$, and the scattered electric field, as

$$\mathbf{E}_{\mathrm{tot}}(\mathbf{r}) = \mathbf{E}_{\mathrm{inc}}(\mathbf{r}) + \frac{\alpha k^2}{\varepsilon_0}\int_V \overleftrightarrow{\mathbf{G}}(\mathbf{r},\mathbf{r}')\mathbf{E}_{\mathrm{inc}}(\mathbf{r}')\mathrm{d}V, \tag{1}$$

where $\mathbf{r}$ is the observation point, $\mathbf{r}'$ is the position of the scattering point, $\alpha$ is the material polarisability, $k$ is the wave vector, $\varepsilon_0$ is the permittivity of free space, and $\overleftrightarrow{\mathbf{G}}(\mathbf{r},\mathbf{r}')$ is the dyadic Green's tensor[28]. This tensor describes how the electric field scatters from an infinitesimal defect on the surface. Scattering occurs via two distinct processes: radiation remnants, which propagate perpendicular to the electric field oscillation direction and form lines parallel to the polarisation; and SPP excitation, which propagates along the polarisation direction and forms NLL lines perpendicular to the laser polarisation. Both effects are included in $\overleftrightarrow{\mathbf{G}}(\mathbf{r},\mathbf{r}')$. Since the average height of the surface structures is significantly smaller than the wavelength of the laser, the three-dimensional volume integral can be reduced to a two-dimensional surface integral,

$$\mathbf{E}_{\mathrm{tot}}(\mathbf{r}) = \mathbf{E}_{\mathrm{inc}}(\mathbf{r}) + \frac{\alpha k^2}{\varepsilon_0}\int_S \overleftrightarrow{\mathbf{G}}(\mathbf{r},\mathbf{r}')\mathbf{E}_{\mathrm{inc}}(\mathbf{r}')h(\mathbf{r}')\mathrm{d}S, \tag{2}$$

where $h(\mathbf{r}')$ is the surface height profile. $\overleftrightarrow{\mathbf{G}}(\mathbf{r},0)$ for Cartesian coordinates (*i.e.*, $\mathbf{r} = r\sin(\theta)\hat{a}_{\mathrm{x}} + r\cos(\theta)\hat{a}_{\mathrm{y}}$) is given by:

$$\overleftrightarrow{\mathbf{G}}(\mathbf{r},0) = e^{-jkr}\begin{bmatrix} \cos^2\theta\frac{1}{r} + [3\sin^2\theta - 1](\frac{1}{k^2r^3} + \frac{j}{kr^2}) & -\sin\theta\cos\theta\frac{1}{r} + 3\sin\theta\cos\theta(\frac{1}{k^2r^3} + \frac{j}{kr^2}) \\ -\sin\theta\cos\theta\frac{1}{r} + 3\sin\theta\cos\theta(\frac{1}{k^2r^3} + \frac{j}{kr^2}) & \sin^2\theta\frac{1}{r} + [3\cos^2\theta - 1](\frac{1}{k^2r^3} + \frac{j}{kr^2}) \end{bmatrix}. \tag{3}$$

Next, all components of the scattered electric field, $\mathbf{E}_{\mathrm{scattered}}(\mathbf{r})$, in Cartesian coordinates can be written as:

$$E(x,y)_x^x = \frac{\alpha k^2}{\varepsilon_0}\iint E_{\mathrm{inc},x}(x',y')h(x',y')e^{-jkr}\left[(3\sin^2\theta - 1)(\frac{1}{k^2r^3} + \frac{j}{kr^2}) + \frac{\cos^2\theta}{r}\right]\mathrm{d}x'\mathrm{d}y', \tag{4}$$

$$E(x,y)_x^y = \frac{\alpha k^2}{\varepsilon_0}\iint E_{\mathrm{inc},x}(x',y')h(x',y')e^{-jkr}\left[3\sin\theta\cos\theta(\frac{1}{k^2r^3} + \frac{j}{kr^2}) - \frac{\sin\theta\cos\theta}{r}\right]\mathrm{d}x'\mathrm{d}y', \tag{5}$$

$$E(x,y)_y^y = \frac{\alpha k^2}{\varepsilon_0}\iint E_{\mathrm{inc},y}(x',y')h(x',y')e^{-jkr}\left[(3\cos^2\theta-1)\left(\frac{1}{k^2r^3}+\frac{j}{kr^2}\right)+\frac{\sin^2\theta}{r}\right]\mathrm{d}x'\mathrm{d}y', \tag{6}$$

$$E(x,y)_y^x = \frac{\alpha k^2}{\varepsilon_0}\iint E_{\mathrm{inc},y}(x',y')h(x',y')e^{-jkr}\left[3\sin\theta\cos\theta(\frac{1}{k^2r^3}+\frac{j}{kr^2})-\frac{\sin\theta\cos\theta}{r}\right]\mathrm{d}x'\mathrm{d}y', \tag{7}$$

where $E(x,y)_x^x$ is the electric field in the $x$-direction induced by an $x$-polarised laser beam, $E(x,y)_x^y$ is the electric field in the $y$-direction caused by an $x$-polarised laser beam. Here, $E_{\mathrm{inc},x}$ and $E_{\mathrm{inc},y}$ are the incident electric fields along the $x$ and $y$ directions, respectively. The angular variable is defined as $\theta = \arctan[(y-y')/(x-x')]$, and the radial distance is $r=\sqrt{(x-x')^2+(y-y')^2}$.

Since a full quantitative description of surface plasmon polariton (SPP) excitation requires integration over all angles near the defect, it is quite complex. For our purposes, we simplify this by introducing a coupling constant $\kappa$, which represents the coupling between the near-field electric field and the SPP. As SPP waves decay exponentially with distance from the source[29], the resulting equations become:

$$E(x,y)_x^x = \frac{\alpha k^2}{\varepsilon_0}\iint E_{0\mathrm{x}}(x',y')h(x',y')e^{-jkr}\left[\kappa(3\sin^2\theta-1)e^{-\beta r}+\frac{\cos^2\theta}{r}\right]\mathrm{d}x'\mathrm{d}y', \tag{8}$$

$$E(x,y)_x^y = \frac{\alpha k^2}{\varepsilon_0}\iint E_{0\mathrm{x}}(x',y')h(x',y')e^{-jkr}\left(3\kappa\sin\theta\cos\theta e^{-\beta r}-\frac{\sin\theta\cos\theta}{r}\right)\mathrm{d}x'\mathrm{d}y', \tag{9}$$

$$E(x,y)_y^y = \frac{\alpha k^2}{\varepsilon_0}\iint E_{0\mathrm{y}}(x',y')h(x',y')e^{-jkr}\left[\kappa(3\cos^2\theta-1)e^{-\beta r}+\frac{\sin^2\theta}{r}\right]\mathrm{d}x'\mathrm{d}y', \tag{10}$$

$$E(x,y)_y^x = \frac{\alpha k^2}{\varepsilon_0}\iint E_{0\mathrm{y}}(x',y')h(x',y')e^{-jkr}\left(3\kappa\sin\theta\cos\theta e^{-\beta r}-\frac{\sin\theta\cos\theta}{r}\right)\mathrm{d}x'\mathrm{d}y', \tag{11}$$

where $(2\beta)^{-1}$ is the SPP decay constant, leading to $\mathrm{NLL}_\perp$. The terms that exclude $\kappa$ correspond to the dipole-like (for linearly polarised laser light) radiation responsible for $\mathrm{NLL}_\parallel$.

Finally, the total intensity on the material surface is given by:

$$I_{\mathrm{tot}}(x,y) = |E(x,y)_x^x\hat{a}_x + E(x,y)_y^x\hat{a}_x + E(x,y)_x^y\hat{a}_y + E(x,y)_y^y\hat{a}_y|^2, \tag{12}$$

where $\hat{a}_x$ and $\hat{a}_y$ are unit vectors along the $x$ and $y$ directions, respectively.

The scattering kernel, $K$, is a reduction of Eqn. (8) - (11) to scalar form. For instance, the incident electric field vector for a $x$-polarised laser beam will be $\mathbf{E}_{\mathrm{inc}}(x,y) = E_{\mathrm{inc},x}(x,y)\hat{a}_x$. All the components including $E_{\mathrm{inc},y}(x,y)$ as a multiplicative term, namely Eqns. (10), (11), vanish. Eqn. (9) remains complex only and does not contribute to interference in Eqn. (12) but contributes a bias to the overall interference pattern. Therefore, the total intensity can be written as:

$$I_{\mathrm{tot}}(x,y) = \left|E_{\mathrm{inc},x}(x,y) + \frac{\alpha k^2}{\varepsilon_0}\iint E_{\mathrm{inc},x}(x',y')h(x',y')K(x-x',y-y')\mathrm{d}x'\mathrm{d}y'\right|^2, \tag{13}$$

with a kernel, $K$, of:

$$K(x-x', y-y') = e^{-jkr}\left[\kappa(3\cos^2\theta - 1)e^{-\beta r} + \frac{\sin^2\theta}{r}\right]. \tag{14}$$

We calculate the total intensity distribution by considering contributions from every infinitesimal point along the surface, where each point acts as an individual scatterer with strength proportional to its local height profile, as expressed in Eq. (2).

A single defect is modelled by setting $h(x, y) = \delta(x, y)$, where $\delta(x, y)$ is the Dirac delta function. The near-field component of the SPP creates a short-range intensity modulation perpendicular to the polarisation axis (green curve). This behaviour is expected, as SPP excitation decays rapidly (within a few micrometres) for high-loss metals such as Ti (SSP decay length of 4.6 μm), Cr, and W. In contrast, SPPs can propagate several tens of micrometres in low-loss metals, such as Au, Ag, and Cu.

The resulting intensity distribution reveals two key features:

(i) A high-intensity, short-range modulation perpendicular to the laser polarisation corresponds to $NLL_{\perp}$.

(ii) A lower-intensity, long-range modulation parallel to laser polarisation corresponds to $NLL_{\parallel}$.

Now we describe the modelling of surface modifications through ablation and thermochemical reactions. To calculate the volume of oxidised (in ambient air) and ablated material under the laser beam's influence, we use the following expressions:

$$\mathrm{d}V_{\mathrm{ox}}(x, y) = \mathrm{d}A \int f_{\mathrm{ox}}(x, y, z)\mathrm{d}z, \tag{15}$$

$$\mathrm{d}V_{\mathrm{ab}}(x, y) = \mathrm{d}A \int f_{\mathrm{ab}}(x, y, z)\mathrm{d}z, \tag{16}$$

where $\mathrm{d}V_{\mathrm{ox}}(x, y)$ and $\mathrm{d}V_{\mathrm{ab}}(x, y)$ are the activated volume elements, $f_{\mathrm{ox}}(x, y, z)$ and $f_{\mathrm{ab}}(x, y, z)$ are the activation probability functions for oxidation and ablation, respectively. Both processes are modelled by assuming effective thresholds, below which no reaction occurs. The oxidation threshold is calculated as the energy required to convert the surface material from the solid phase at room temperature to the activated state (near melting but below evaporation). For cobalt atoms, we use the following material properties: the specific heat capacity, $c_{\mathrm{sp}} = 0.42$ J/gK, fusion heat, $H_{\mathrm{f}} = 16.19$ kJ/mol, melting temperature, $T_{\mathrm{m}} = 1768$ K, boiling temperature, $T_{\mathrm{b}} = 3200$ K, density, $\rho = 8.9$ g/cm$^3$, atomic mass, $M = 58.93$ amu, skin depth, $z_{\mathrm{s}} = 29$ nm, and pulse duration, $\tau_{\mathrm{p}} = 300$ fs. Similar calculations have been made for all other materials used in this study.

The ablation threshold is calculated assuming that if the vaporisation energy is exceeded, the material will be ablated and removed from the material surface. For cobalt, the heat of vaporisation is $H_{\mathrm{v}} = 376.5$ kJ/mol. With these coefficients and assumptions, the oxidation and ablation thresholds are calculated as $I_{\mathrm{ox}} = 110$ GW/cm$^2$, and $I_{\mathrm{ab}} = 600$ GW/cm$^2$, respectively. These threshold calculations are simplified approximations of highly complex processes. Nevertheless, they align well with experimental observations. During simulations, we evaluate both thresholds at each surface point after every pulse to determine which process (if any) is activated.

Since real threshold processes are not discontinuous, directly imposing sharp thresholds causes numerical complications. To address this, we introduce smoothened threshold functions using differentiable sigmoid-like functions such as:

$$f_{\rm ox}(x,y) = \frac{1+\mathrm{erf}\{s_{\rm ox}[I_{\rm tot}(x,y) - I_{\rm ox}]\}}{2}, \tag{17}$$

where $\mathrm{erf}\{\cdot\}$ is the error function, and $s_{\rm ox}$ is a constant that sets the sharpness of the thresholds. The expression for ablation follows a similar form and is omitted here for brevity. Both activation probability functions are plotted in Figure. S1. The incident pulse intensity at depth $z$ within the material is given by:

$$I_{\rm tot}(x,y,z) = I_{\rm surf,tot}(x,y)e^{-z/z_{\rm s}}, \tag{18}$$

where $I_{\rm surf,tot}(x,y)$ is the intensity distribution, and $z_{\rm s}$ is the skin depth of the material.

Finally, the corresponding volume elements for oxidation and ablation are expressed as follows:

$$\mathrm{d}V_{\rm ox}(x,y) = \mathrm{d}A\, z_{\rm s} c_{\rm ox}\big(\mathrm{erf}\{\mathrm{s}_{\rm ox}[I_{\rm tot}(x,y) - I_{\rm ox}]\} + 1\big)/2, \tag{19}$$

$$\mathrm{d}V_{\rm ab}(x,y) = \mathrm{d}A\, z_{\rm s} c_{\rm ab}\big(\mathrm{erf}\{\mathrm{s}_{\rm ab}[I_{\rm tot}(x,y) - I_{\rm ab}]\} + 1\big)/2 \tag{20}$$

where $c_{\rm ox}$ and $c_{\rm ab}$ are proportionality factors setting the rates of oxidation and ablation processes, respectively.

To calculate the change in surface height per laser pulse due to oxidation, we must account for both the number of activated surface atoms and the availability of $O_2$ molecules in the immediate vicinity.

When a laser pulse arrives, the number of, *e.g.*, Co atoms activated by that pulse within the volume, $\mathrm{d}V_{\rm ox}(x,y)$, is given by, $N_{\rm Co,ox} = \mathrm{d}V_{\rm ox}(x,y)\rho N_{\rm A}/M_{\rm Co}$, where $M_{\rm Co}$ is the molar mass of cobalt, $\rho$ is its mass density, $N_{\rm A}$ is Avogadro's number.

The number of available oxygen molecules for the reaction is given by $N_{\rm O_2}(x,y) = c_{\rm O_2} e^{-h_n(x,y)/h_{\rm c}}$, where $h_n(x,y)$ (if positive) is the oxide layer thickness after the $n^{\rm th}$ pulse and $h_{\rm c}$ is the critical height that characterises oxygen diffusion into the cobalt oxide layer.

Following each pulse, the corresponding change in the oxide layer is given by, $\Delta h_{{\rm ox},n}(x,y) = r_{\rm ox} N_{\rm CoO}$, where $r_{\rm ox}$ is the oxidation rate and $N_{\rm CoO}$ is the number of newly formed oxide molecules. The latter is determined by the limiting factor in the reaction,

$$N_{\rm CoO} = \begin{cases} N_{\rm Co,ox} & \text{if } N_{\rm Co,ox} < N_{\rm O_2}, \\ N_{\rm O_2} & \text{if } N_{\rm O_2} < N_{\rm Co,ox}\,. \end{cases}. \tag{21}$$

This self-limiting nature of the oxidation process introduces a form of negative feedback in NLL, which naturally saturates the growth of surface structures.

The height change due to the ablation process can be calculated in a similar manner. The activated volume element, $\mathrm{d}V_{\mathrm{ab}}(x, y)$, is determined similarly, using the parameters $I_{\mathrm{ab}}$, $s_{\mathrm{ab}}$, and $c_{\mathrm{ab}}$. However, unlike oxidation, the ablated material is simply ejected without involving any additional reactants. The corresponding height change is given by $\Delta h_{\mathrm{ab},n}(x, y) = \mathrm{d}V_{\mathrm{ab}}(x, y)/\mathrm{d}A = -r_{\mathrm{ab}}N_{\mathrm{Co,ab}}$, which is always a negative function, except when zero. Here, $r_{\mathrm{ab}}$ is the ablation rate and $N_{\mathrm{Co,ab}}$ is the number of surface molecules activated for ablation.

Combining both mechanisms, the net height change after the $n^{\mathrm{th}}$ pulse is given by

$$\Delta h_n(x, y) = \Delta h_{\mathrm{ox},n}(x, y) + \Delta h_{\mathrm{ab},n}(x, y). \tag{22}$$

Before proceeding with the calculations for the $(n+1)^{\mathrm{st}}$ pulse, the height profile is updated as,

$$h_{n+1}(x, y) = h_n(x, y) + \Delta h_n(x, y). \tag{23}$$

**Numerical simulations**

The simulation code runs over the entire cycle (Figure 1d), with the position of the focal spot moved according to the scanning speed. It is important to note that the beam motion is much slower than the pulse repetition rate and the associated physical processes.

The numerical solutions of our theoretical model have consistently matched experimental results with high accuracy. As the first demonstration, we show the ability to create both types of NLL structures on the same surface and to switch between them (Figure S4 and Videos 3-5). While similar calculations have been performed for all the materials considered in this study, for this example, we use the parameters for Ti: oxidation threshold $I_{\mathrm{ox}} = 0.184$ TW/cm², ablation threshold $I_{\mathrm{ox}} = 0.902$ TW/cm², surface plasmon decay length $\beta = 4.61$ µm, surface plasmon coupling coefficient $\kappa = 10^5$, oxidation rate $r_{\mathrm{ox}} = 10^{-3}$, ablation rate $r_{\mathrm{ab}} = 5 \times 10^{-16}$, wave number $k = 2\pi/\Lambda$ with $\Lambda = 870$ nm, critical thickness for oxidation $h_{\mathrm{c}} = 15$ nm, and $\alpha k^2/\varepsilon_0 = 5 \times 10^{12}$. In the simulations, we raster-scan the laser beam starting from the top-left corner of the field of view, initially using the laser parameters corresponding to $\mathrm{NLL}_{\parallel}$. In the middle of the simulation, we momentarily stop the pulses and change the laser parameters to correspond to $\mathrm{NLL}_{\perp}$, while keeping the laser polarisation the same.

**Calculating the periods of the NLL patterns**

The structure period is determined by the period of the scattering event and the interference over the material surface. In case of radiation remnants ($\mathrm{NLL}_{\parallel}$) the electric field component is concentrated along the surface. The period of the structures can be crudely approximated by $\lambda_0/n_{\mathrm{eff}}$, where $\lambda_0$ is the laser wavelength in free space and $n_{\mathrm{eff}}$ is the effective refractive index, determined by a combination of the indices of ambient air, metal (or semiconductor), the corresponding oxide material, and the amount (thickness) of the oxide. We assumed $n_{\mathrm{eff}} = a n_{\mathrm{air}} + b n_{\mathrm{material}} + c n_{\mathrm{oxide}}$. Since retarded waves travel dominantly through ambient air, $a = 0.75$, $b = c = 0.125$, in most cases. In the case of Ni, we assumed $a_{\mathrm{Ni}} = 0.9$, $b_{\mathrm{Ni}} = 0.02$, $c_{\mathrm{Ni}} = 0.08$, and for Nb, we assumed $a_{\mathrm{Nb}} = 0.6$, $b_{\mathrm{Nb}} = 0.04$, $c_{\mathrm{Nb}} = 0.36$. If the formation is due to SPP coupling to the material ($\mathrm{NLL}_{\perp}$), the structure period is given by $\lambda = \kappa 2\pi / \mathrm{Re}\{k_{\mathrm{SPP}}\}$, where $\kappa$ is a scaling constant, taken as 0.75 here, $k_{\mathrm{SPP}} = k_0 \sqrt{\varepsilon_{\mathrm{metal}} \varepsilon_{\mathrm{air}} / (\varepsilon_{\mathrm{metal}} + \varepsilon_{\mathrm{air}})}$, with $k_0 = 2\pi/\lambda_0$, and $\varepsilon_{\mathrm{metal,air}}$ the relative permittivities of the metal and air, respectively.

A special case is silicon. Because it is a semiconductor, it must be excited to reveal its metallic property and to support SPP waves. Further, its refractive index scales with the excitation strength. Given that the laser pulses are femtoseconds-long, there are several mechanisms, such as multi-photon absorption, free carrier absorption, and Auger recombination. Due to the complexity of these processes, we used empirical observations to estimate the period in Si[39] instead of a direct calculation like we did for the other materials. Figure 3b illustrates the estimation of the structure period and the measured structure period. The sample tilt causes an experimental uncertainty in the setup with respect to the incidence angle of the beam, which is expressed by the error bar, corresponding to a 0.5-degree uncertainty for the tilt over the sample. Since the refractive indices and permittivities vary wildly in the literature, it is not possible to estimate the structure period precisely. Therefore, we assumed a 30% uncertainty on the refractive indices of the metal and metal-oxides. We used the permittivities (refractive indices) given in Table 2. Despite these uncertainties, the estimated periods agree reasonably well with the experimental measurements of the period structures over all the materials considered in this study.

The relationship between the laser structure period and the incidence angle is given by

$$\Lambda = \frac{\Lambda_0}{1 \pm \sin\theta}, \tag{24}$$

where $\Lambda$ is the resulting period, $\Lambda_0$ is the nominal period, and $\theta$ is the incidence angle with respect to the surface normal[11]. Figure S5 shows the comparison of the theoretical, experimental and simulation results for a Ti surface. The black straight line is the period estimated by Eqn. (24).

**Table 2.** Optical constants of the processed materials.

| Material | $\varepsilon_r$ (real permittivity) | $\varepsilon_i$ (imaginary permittivity) | Bulk/ Thin film | Reference | Material | $\varepsilon_r$ (real permittivity) | $n$ (refractive index) | Bulk/Thin film | Reference |
|---|---|---|---|---|---|---|---|---|---|
| Co | -24,33 | 32,7 | Bulk | [30] | CoO | 4,6656 | 2,16 | Bulk | [31] |
| Nb | -25,35 | 16,95 | Bulk | [32] | $Nb_2O_5$ | 5,0176 | 2,24 | Thin film | [33] |
| Ni | -26,5 | 29,6 | Bulk | [30] | NiO | 3,24 | 1,8 | Bulk | [34] |
| Cr | -0,52 | 25,04 | Thin film | [30] | $Cr_2O_3$ | 3,61 | 1,9 | Thin film | [35] |
| Al | -97,44 | 27,81 | Bulk | [36] | $Al_2O_3$ | 3,0625 | 1,75 | Bulk | [37] |
| Ti | -4,2 | 27,43 | Bulk | [30] | $TiO_2$ | 6,1504 | 2,48 | Bulk | [38] |
| Si | N/A (see text) | N/A (see text) | Bulk | [39] | $SiO_2$ | 2,1025 | 1,45 | Bulk | [40] |

## Figure Captions

**Figure 1 Competing pattern forming mechanisms.** Sketches of structure formation **(a)** parallel (oxidation-type $\mathrm{NLL}_{\parallel}$) and **(b)** perpendicular (ablation-type $\mathrm{NLL}_{\perp}$) to the laser polarisation. The laser beam is depicted in red. Pattern formation is colour-coded: oxidation-type patterns are shown in blue, while ablation-type patterns are represented in green. **(c)** Numerical simulations showing **(i)** the initial surface profile ($h_{n=1}$) that is excited by the first laser pulse where the blue dot represents the strongest scatterer on the surface, and **(ii)** the total intensity distribution ($I_{\mathrm{total},n=1}$) after scattering. The scattering mechanisms responsible for $\mathrm{NLL}_{\parallel}$ and $\mathrm{NLL}_{\perp}$ structures are simultaneously present, which are easily recognised from the **(iii)** vertical (green) and horizontal (blue) cross-sections of $I_{\mathrm{total},1}$. **(d)** Flowchart summarising the theory of feedback-driven NLL process involving the surface profile, $h_n$, electric field profile, $E_n$, scattering kernel, $K$, and total intensity profile, $I_{\mathrm{total},n}$ for the $n^{\mathrm{th}}$ incident pulse. **(e)** Numerical simulations show that, depending on the pulse fluence, one mechanism can dominate while completely suppressing the other. The initial surface is shown at $t_0$ with the contours showing the scattered intensity profile induced by the coexisting $\mathrm{NLL}_{\parallel}$ and $\mathrm{NLL}_{\perp}$ (leftmost). $\mathrm{NLL}_{\parallel}$ (upper row) can be selectively activated with the choice of a pulse fluence of, *e.g.*, 26.5 mJ/cm$^2$, whereas 160 mJ/cm$^2$ will favour $\mathrm{NLL}_{\perp}$ (lower row) formation, either resulting in the complete exclusion of the other. The times representing $t_1$ = 97,000$^{\mathrm{th}}$, $t_2$ = 98,000$^{\mathrm{th}}$, $t_3$ = 99,000$^{\mathrm{th}}$, and $t_4$ = 100,000$^{\mathrm{th}}$ pulses for $\mathrm{NLL}_{\parallel}$ and $t_1$ = 4$^{\mathrm{th}}$, $t_2$ = 5$^{\mathrm{th}}$, $t_3$ = 7$^{\mathrm{th}}$, and $t_4$ = 8$^{\mathrm{th}}$ pulses for $\mathrm{NLL}_{\perp}$. The SEM images (rightmost) showing the empirical validation of the simulations on a Ti surface.

**Figure 2 Demonstration of the universality.** The optical images of **(a)** $\mathrm{NLL}_{\parallel}$ (upper row) and $\mathrm{NLL}_{\perp}$ (lower row) patterns on niobium, chromium, titanium, cobalt, nickel, aluminium, and silicon. Corresponding reciprocal peaks (red in white background) indicating high periodicity for both $\mathrm{NLL}_{\parallel}$ and $\mathrm{NLL}_{\perp}$. All images were captured with a 60× objective, with the laser polarisation direction indicated in each row. Scale bars are 5 µm. **(b)** Cross-sectional SEM images of **(i)** $\mathrm{NLL}_{\parallel}$ and **(ii)** $\mathrm{NLL}_{\perp}$ on 400-nm-thick Ti, and **(iii)** $\mathrm{NLL}_{\parallel}$ and **(iv)** $\mathrm{NLL}_{\perp}$ on 100-nm-thick cobalt surfaces showing $\mathrm{NLL}_{\parallel}$ patterns are growing on the surface (red dashed lines) as oxide structures, whereas $\mathrm{NLL}_{\perp}$ patterns are formed as depressions into the bulk material. The scanning speed was 1.5 m/s, with a spot size of ~10 µm at a repetition rate of 1-MHz and 350 mW of average power, corresponding to ~6 pulses per spot with 3 kJ/m$^2$ pulse fluence. **(c)** The parameter space of NLL formation as a function of pulse fluence and the number of pulses per spot. A low number of high fluence pulses leads to $\mathrm{NLL}_{\perp}$ formation (green), whereas a high number of low fluence pulses favours $\mathrm{NLL}_{\parallel}$ formation (blue). The plot also includes the parameter range where there are no (white) or irregular (beige) modifications. Red crosses denote experimental results

performed on the Co-deposited glass substrate, which are in good agreement with the theoretical calculations. **(d)** Empirically measured (circles) and theoretically calculated (ellipsoids) periods of $NLL_{\parallel}$ (blue circles and pink ellipsoids) and $NLL_{\perp}$ (green circles and beige ellipsoids) structures with their formation mechanisms indicated as subscripts for each of the materials shown in **(a)**.

**Figure 3 Proof-of-concept device demonstrations. (a)** A holographic temperature sensor shown on 110-nm-thick Co-coated Kapton tape. The periodicity ($\Lambda_n$) of the $NLL_{\parallel}$ structures changes with the temperature due to expansion or contraction of the polymer tape. It increases from **(i)** T = 27 °C to **(ii) T** = 100 °C and decreases back when **(iii)** T = 37 °C. **(b)** The device is illuminated with white light, and the red-colour channel of the recorded image is analysed. The change in reflectivity for red colour is used to sense the temperature changes. **(c)** Demonstration of surface wettability control. **(i)** Superhydrophilic and **(ii)** superhydrophobic $NLL_{\perp}$ functionalisation of silicon wafer. **(iii)** Time-lapse images of a water droplet bouncing on an $NLL_{\perp}$ functionalised superhydrophobic surface. **(iv)** SEM image of seamlessly adjacent superhydrophilic and superhydrophobic regions created by $NLL_{\perp}$ structures. **(d)** Structural colouring of a cobalt surface with $NLL_{\parallel}$ and $NLL_{\perp}$ patterns by using two mutually orthogonal polarisation directions and formed by enacting 100 pre-programmed transitions between $NLL_{\parallel}$ and $NLL_{\perp}$. The image was recorded under white light illumination. The $NLL_{\parallel}$ patches were formed using 200,000 pulses/spot, each with a fluence of 42 mJ/cm$^2$, whereas 60 pulses per spot, each with 210 mJ/cm$^2$ created $NLL_{\perp}$ patches. **(e)** A structural-colour recreation of the wings of a *Limenitis arthemis astyanax* butterfly (inset) on a thin-film coated flexible glass, where the perceived colours change, similarly to the real butterfly's wings, when the glass is bent to mimic the wing flapping.

## Video Captions

**Video 1 Empirical demonstration of nonlocal feedback driving $NLL_{\parallel}$ formation.** Initially, patterns emerge on a bare surface, with the pattern edge appearing more pronounced than the centre. This asymmetry results from the beam spending more time stationary at the edges while scanning continuously across the central region. In the subsequent stage, when patterning occurs adjacent to pre-existing structures, nonlocal feedback is enhanced uniformly across the sample. As a result, visible lines form simultaneously and evenly, despite variations in scanning speed between the centre and the edges.

**Video 2 Empirical demonstration of nonlocal feedback driving $NLL_{\parallel}$ formation.** Continuing from Video 1, this video shows continuous laser scanning beginning on a bare surface and progressing into regions with pre-existing structures, illustrating the transition and exponential amplification of the $NLL_{\parallel}$ mechanism due to nonlocal feedback across the sample.

**Video 3 Numerical simulations showing switching between $NLL_{||}$ and $NLL_{\perp}$ forming mechanisms.** Calculations are performed for cobalt. Low pulse fluence at high repetition rate results in $NLL_{||}$ (yellow lines scanned from the lower to the upper part of the sample), and high pulse fluence at low repetition rate creates $NLL_{\perp}$ (blue lines scanned from the upper to the lower part of the sample).

**Video 4 Numerical simulations showing seamless stitching of multiple NLL regions.** Calculations are performed for cobalt. $NLL_{\perp}$ (blue lines) are forming continuously from existing $NLL_{||}$ (yellow lines).

**Video 5 Numerical simulations showing seamless stitching of multiple NLL regions.** Calculations are performed for cobalt. $NLL_{\perp}$ (blue lines) are forming with a 45° angle from existing $NLL_{||}$ (yellow lines).

**Video 6 Proof-of-concept device applications.** Demonstration of the holographic temperature sensor created by NLL-functionalisation on a cobalt-coated Kapton tape.

**Video 7 Proof-of-concept device applications.** Demonstration of superhydrophobicity of NLL-functionalised silicon wafer.

**Video 8 Proof-of-concept device applications.** Demonstration of water flow guidance along an NLL-functionalised superhydrophilic pathway, bordered by NLL-functionalised superhydrophobic regions on a silicon wafer.

**Video 9 Proof-of-concept device applications.** A structural-colour reproduction of the *Limenitis arthemis astyanax* butterfly on thin-film-coated flexible glass. The displayed colours shift with the viewing angle, mimicking the iridescent effect seen in the butterfly's natural wings.

## Acknowledgements

This work received funding from the European Research Council (ERC) projects NLL, PhD, and UniLase under the European Union's Horizon 2020 research and innovation programme (grant agreements No. 617521, No. 853387, and No. 101055055, respectively) and TÜBİTAK (grant agreement No. 120F147).

## Author contributions

Ö.Y., I.P., O.T., S.I., and F.Ö.I designed the research and experiments. I.P. developed the idea and performed the initial experiments. Ö.Y. and A.A. performed the experiments and analysis. S.G. and A.A. supplied the samples and obtained SEM images. P.D. designed and performed the wettability experiments. Ö.Y., S.I., and F.Ö.I. conceived of the theoretical framework. Ö.Y. and Ü.S.N. developed the mathematical model and performed the simulations.

## Data availability

The data supporting the plots within this paper and other findings of this study are available from the corresponding author upon reasonable request.

## Supplementary Information

Supplementary text, Figures S1–S12.

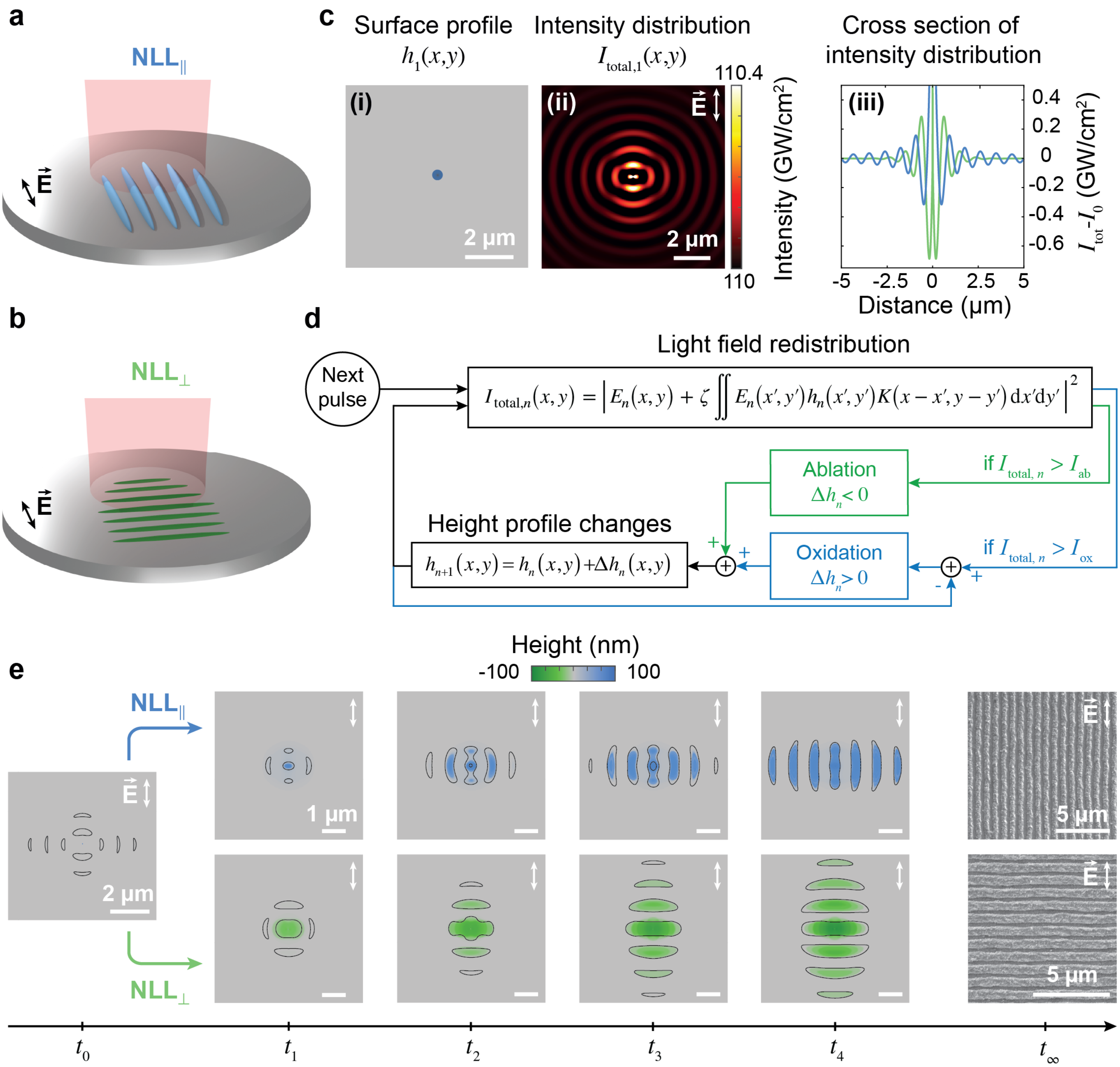

a
NLL∥
E
b
NLL⊥
E
c
Surface profile
$h_1(x,y)$
Intensity distribution
$I_{total,1}(x,y)$
Cross section of
intensity distribution
(i)
2 μm
(ii)
2 μm
110.4
110
Intensity (GW/cm²)
(iii)
0.4
0.2
0
-0.2
-0.4
-0.6
$I_{tot}-I_0$ (GW/cm²)
-5
-2.5
0
2.5
5
Distance (μm)
d
Light field redistribution
Next pulse
$I_{total,n}(x,y) = \left| E_n(x,y) + \zeta \iint E_n(x',y')h_n(x',y')K(x-x',y-y')\,dx'dy' \right|^2$
Ablation
$\Delta h_n < 0$
if $I_{total,n} > I_{ab}$
Oxidation
$\Delta h_n > 0$
if $I_{total,n} > I_{ox}$
Height profile changes
$h_{n+1}(x,y) = h_n(x,y) + \Delta h_n(x,y)$
e
Height (nm)
-100
100
NLL∥
NLL⊥
E
2 μm
1 μm
5 μm
5 μm
$t_0$
$t_1$
$t_2$
$t_3$
$t_4$
$t_\infty$

**Figure 1**

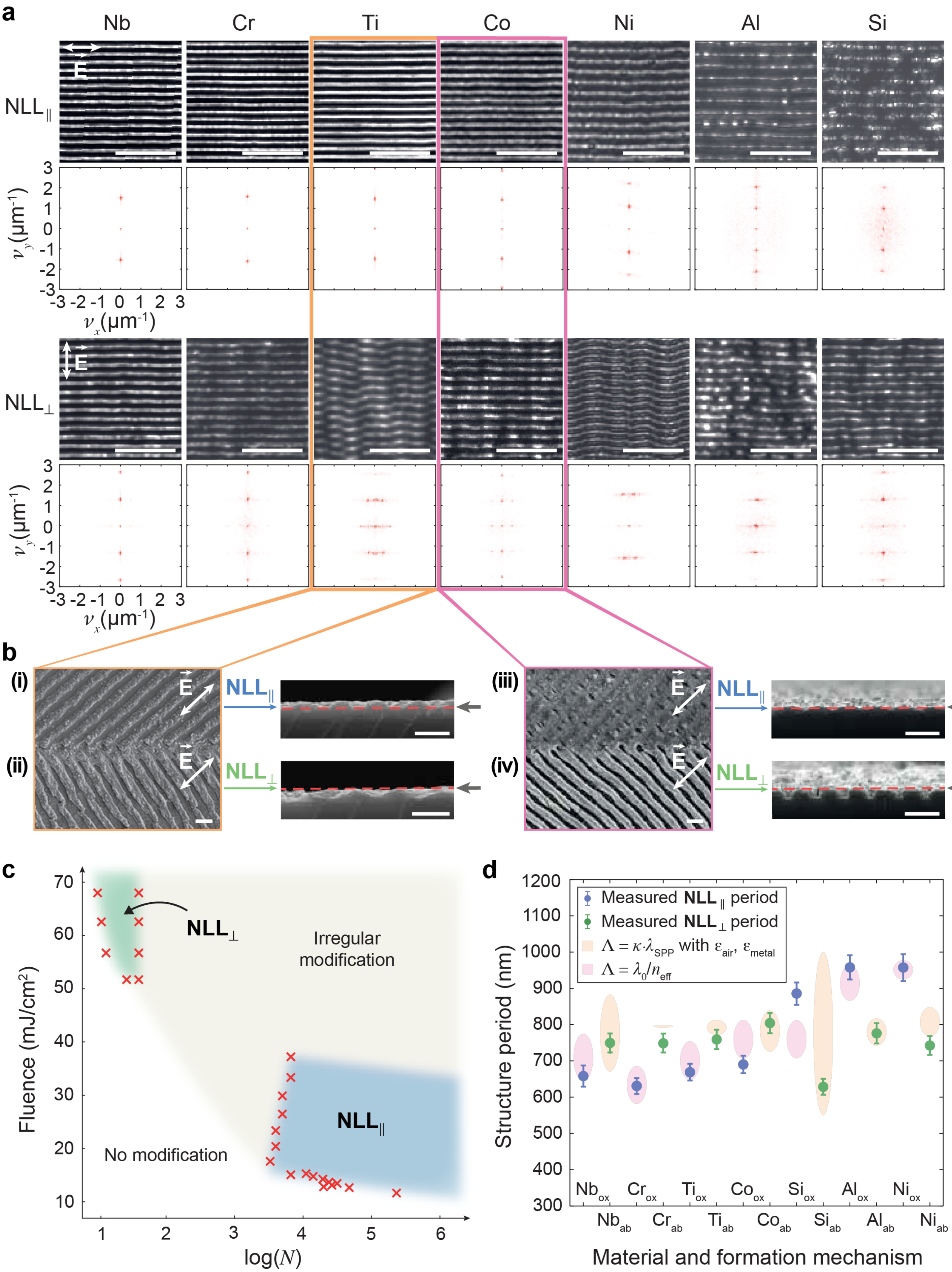

a
Nb
Cr
Ti
Co
Ni
Al
Si
NLL∥
NLL⊥
$\nu_y$(μm$^{-1}$)
$\nu_x$(μm$^{-1}$)
b
(i)
(ii)
(iii)
(iv)
NLL∥
NLL⊥
c
NLL⊥
Irregular modification
NLL∥
No modification
Fluence (mJ/cm$^2$)
log($N$)
d
Measured NLL∥ period
Measured NLL⊥ period
$\Lambda = \kappa \cdot \lambda_{SPP}$ with $\varepsilon_{air}$, $\varepsilon_{metal}$
$\Lambda = \lambda_0 / n_{eff}$
Structure period (nm)
Material and formation mechanism

**Figure 2**

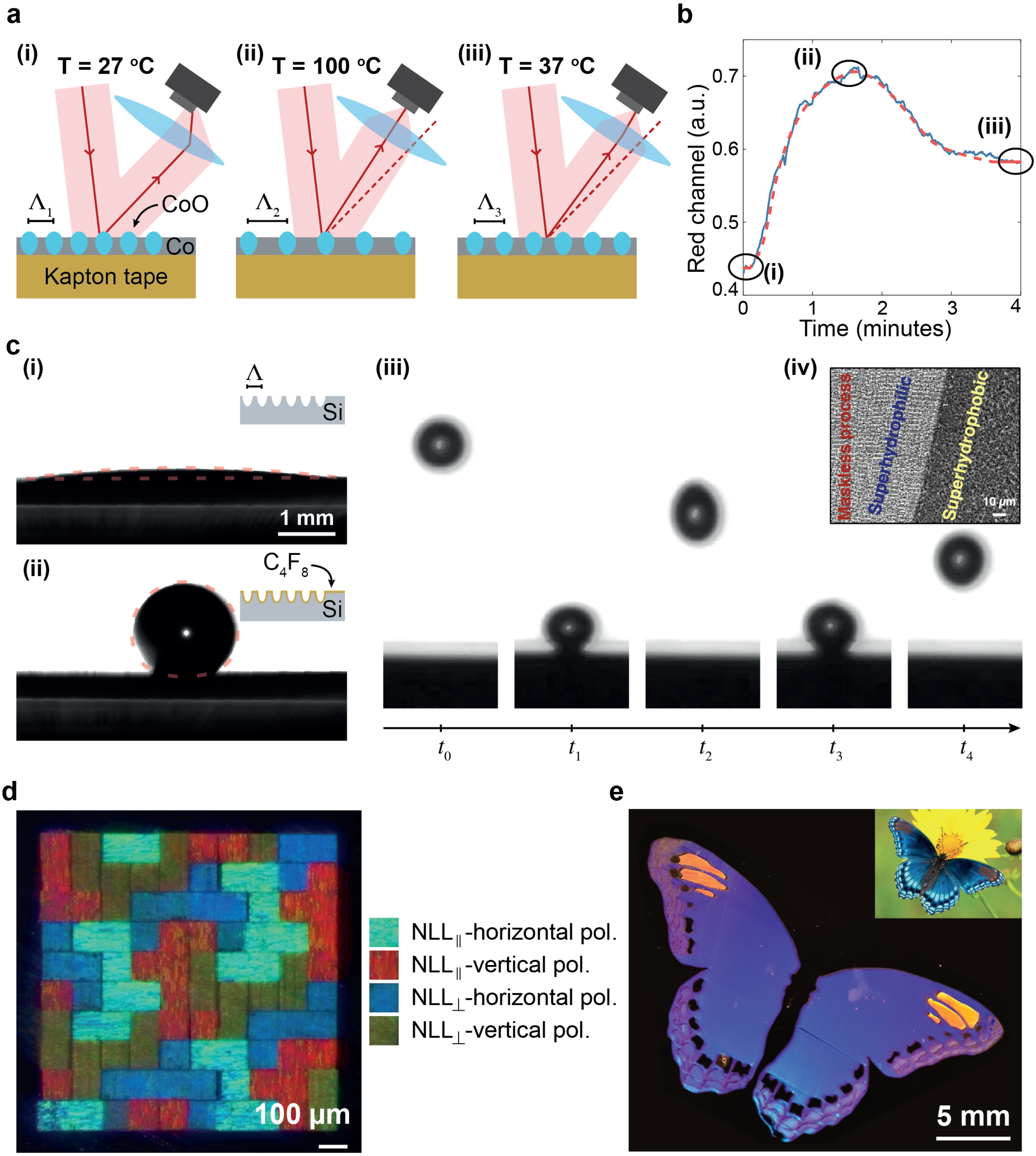
a
(i)
T = 27 °C
(ii)
T = 100 °C
(iii)
T = 37 °C
CoO
Co
Kapton tape
b
Red channel (a.u.)
Time (minutes)
c
(i)
Si
1 mm
(ii)
C4F8
Si
(iii)
(iv)
Maskless process
Superhydrophilic
Superhydrophobic
10 μm
d
NLL∥-horizontal pol.
NLL∥-vertical pol.
NLL⊥-horizontal pol.
NLL⊥-vertical pol.
100 μm
e
5 mm

**Figure 3**

# Controlling Competing Feedback Mechanisms for Programmable Patterns via Nonlinear Laser Lithography

## SUPPLEMENTARY INFORMATION

Özgün Yavuz[1,†], Ihor Pavlov[2], Abdullah bin Aamir[1], Sezin Galioğlu[3], Petro Deminskyi[4], Ü. Seleme Nizam[1], Onur Tokel[3,4], Serim Ilday[1,5], and F. Ömer Ilday[1,5,†]

[1] Faculty of Electrical Engineering and Information Technology, Ruhr Universität Bochum, 44801 Bochum, Germany

[2] Department of Physics, Middle East Technical University, 06800 Ankara, Turkey

[3] UNAM-National Nanotechnology Research Center, Bilkent University, 06800 Ankara, Turkey

[4] Department of Physics, Bilkent University, 06800 Ankara, Turkey

[5] Faculty of Physics and Astronomy, Ruhr Universität Bochum, 44801 Bochum, Germany

[†] Corresponding author: oezguen.yavuz@ruhr-uni-bochum.de & oemer.ilday@ruhr-uni-bochum.de

## Table of Contents

## 1. Thresholds for oxidation and ablation processes

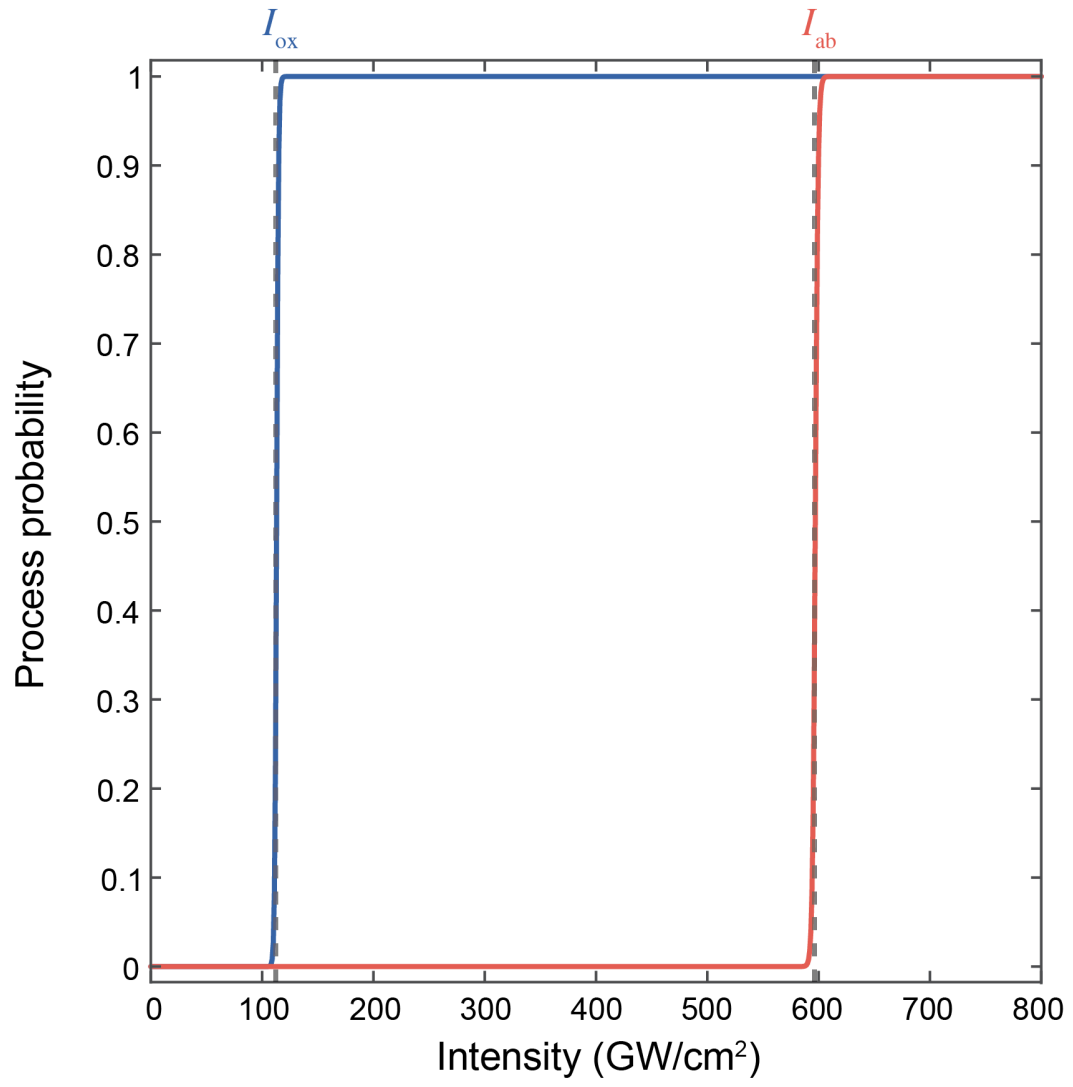

**Figure S1.** The thresholds for oxidation ($I_{ox}$) and ablation ($I_{ab}$) processes are calculated using the theoretical model, described in the Methods section (Eq. 1-18). Both thresholds are described by sigmoid-like functions to avoid discontinuities.

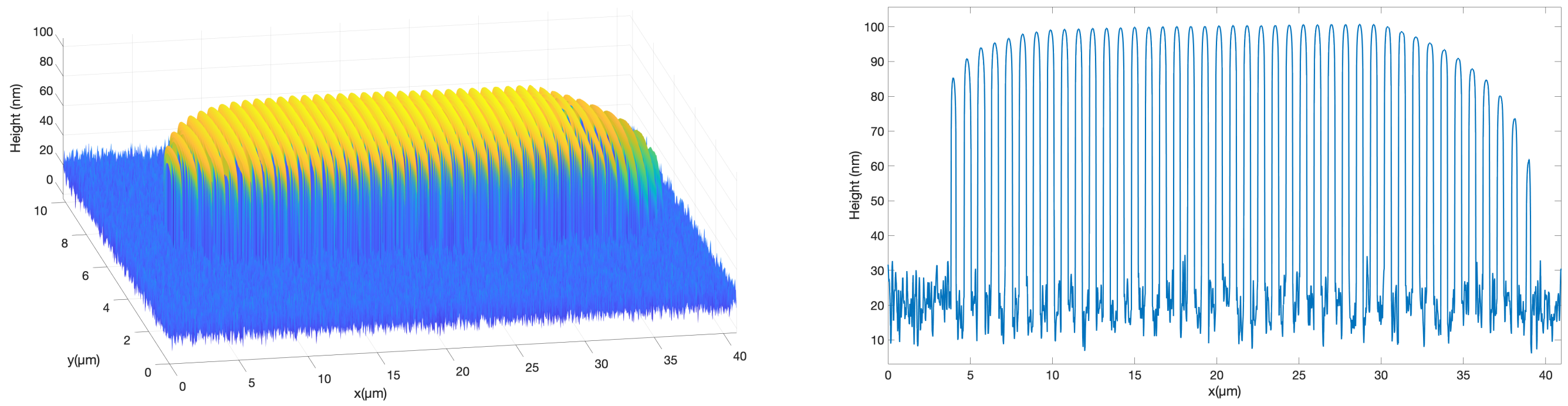

**Figure S2.** Laser scanning microscopy images showing self-similar characteristics of $NLL_{||}$ patterns with uniform height distribution.

## 2. Self-limiting growth as a result of the negative feedback

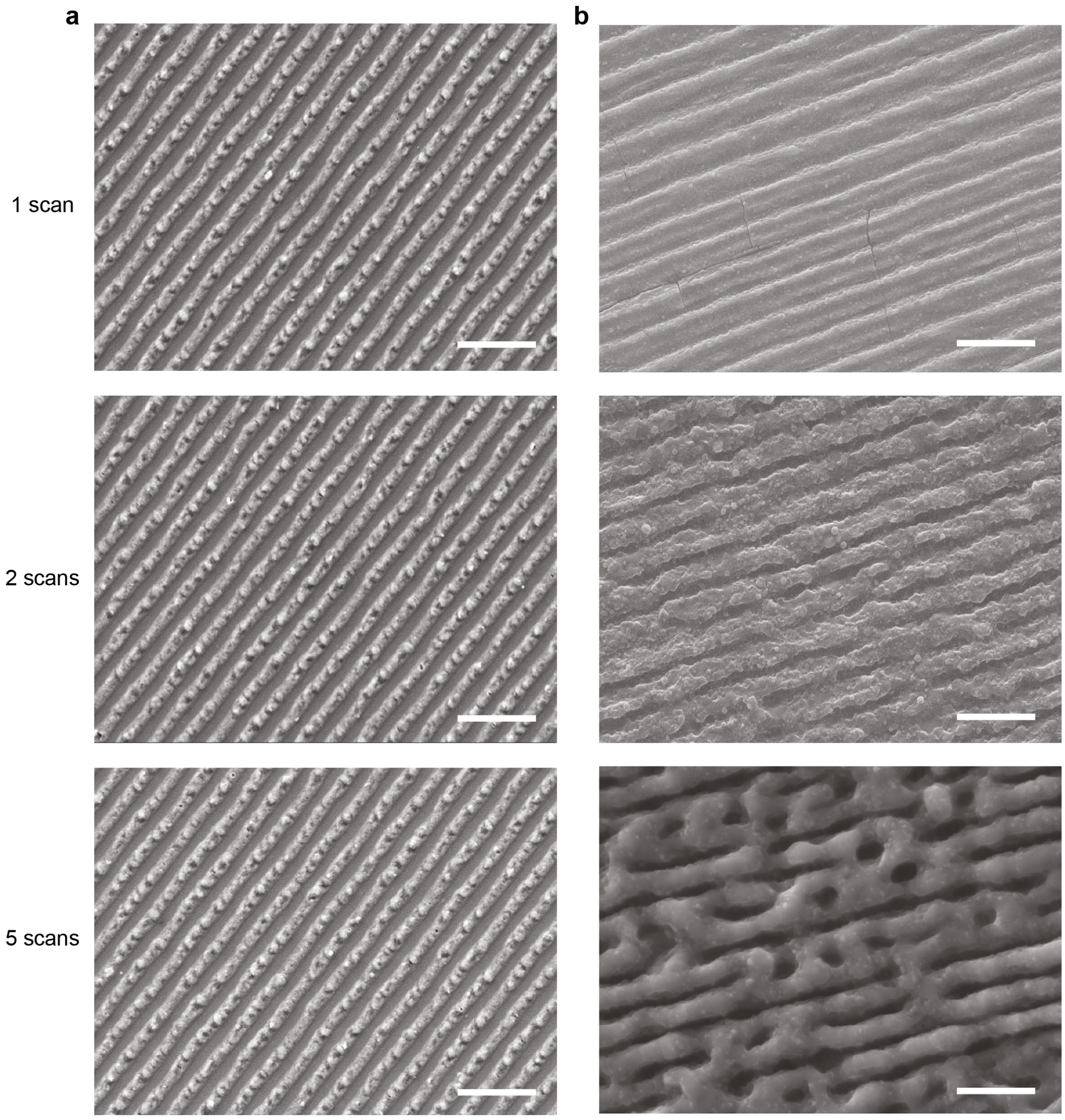

**Figure S3.** Demonstration of the effect of negative feedback on **(a)** $NLL_{\parallel}$ and **(b)** $NLL_{\perp}$ patterns on the titanium surface. Microscopy images showing the surface profile modifications after the first, second, and fifth scans. Scale bars are 2 µm.

## 3. Switching between competing formation mechanisms and tunability of the structure periods

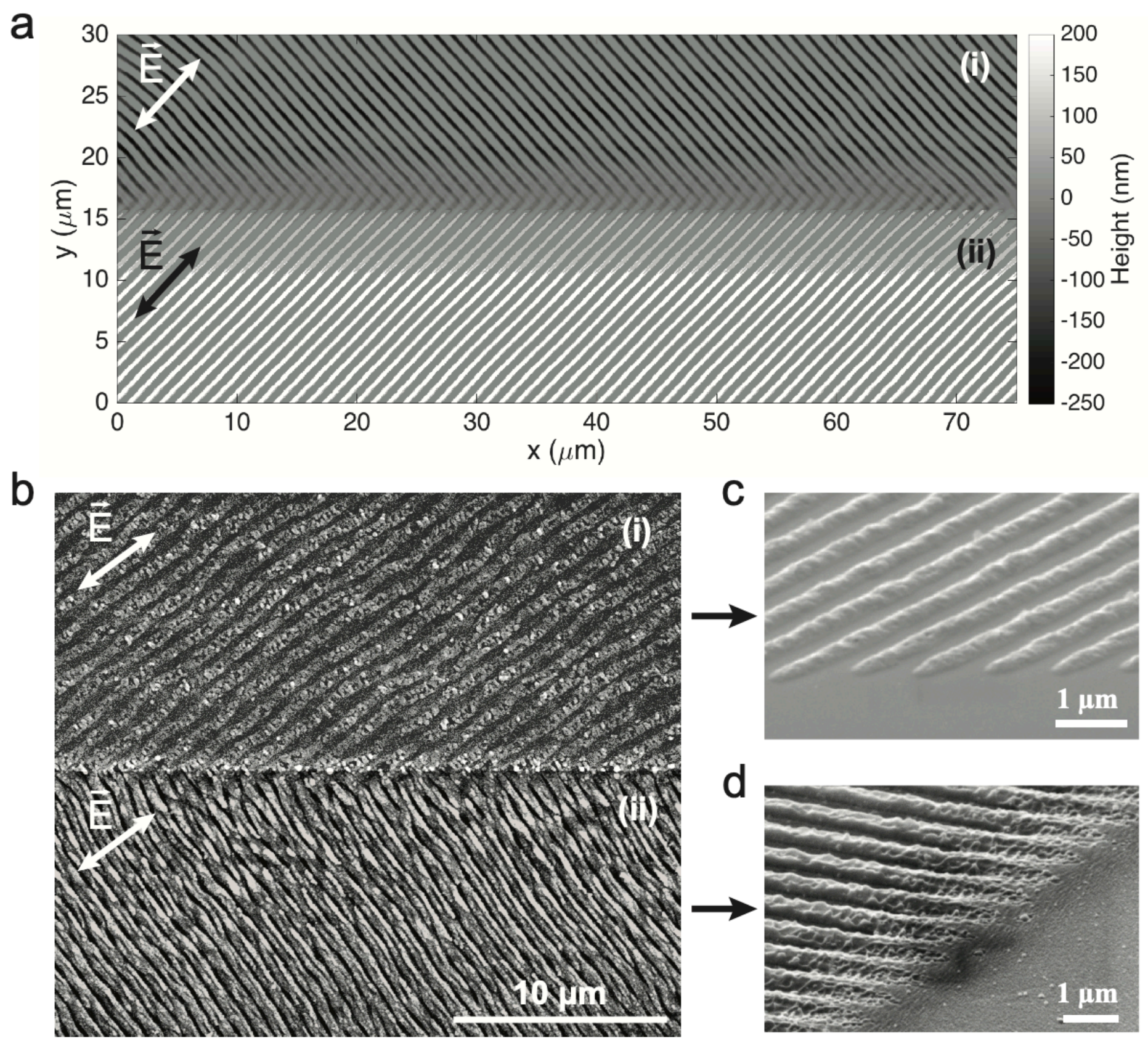

**Figure S4.** Demonstration of the switching mechanism. **(a)** Numerical simulations of a surface scanned with two sets of parameters while the laser polarisation remained the same. At the upper part $I_0 = 0.1$ TW/cm², $f_R = 1$ MHz, which supports oxidation; at the lower part $I_0 = 1.76$ TW/cm², $f_R = 500$ Hz, which supports ablation. Microscopy images showing **(b)** experimental demonstration of the switching between NLL patterns. Cross-sectional view of **(c)** $NLL_{\parallel}$ and **(d)** $NLL_{\perp}$ patterns.

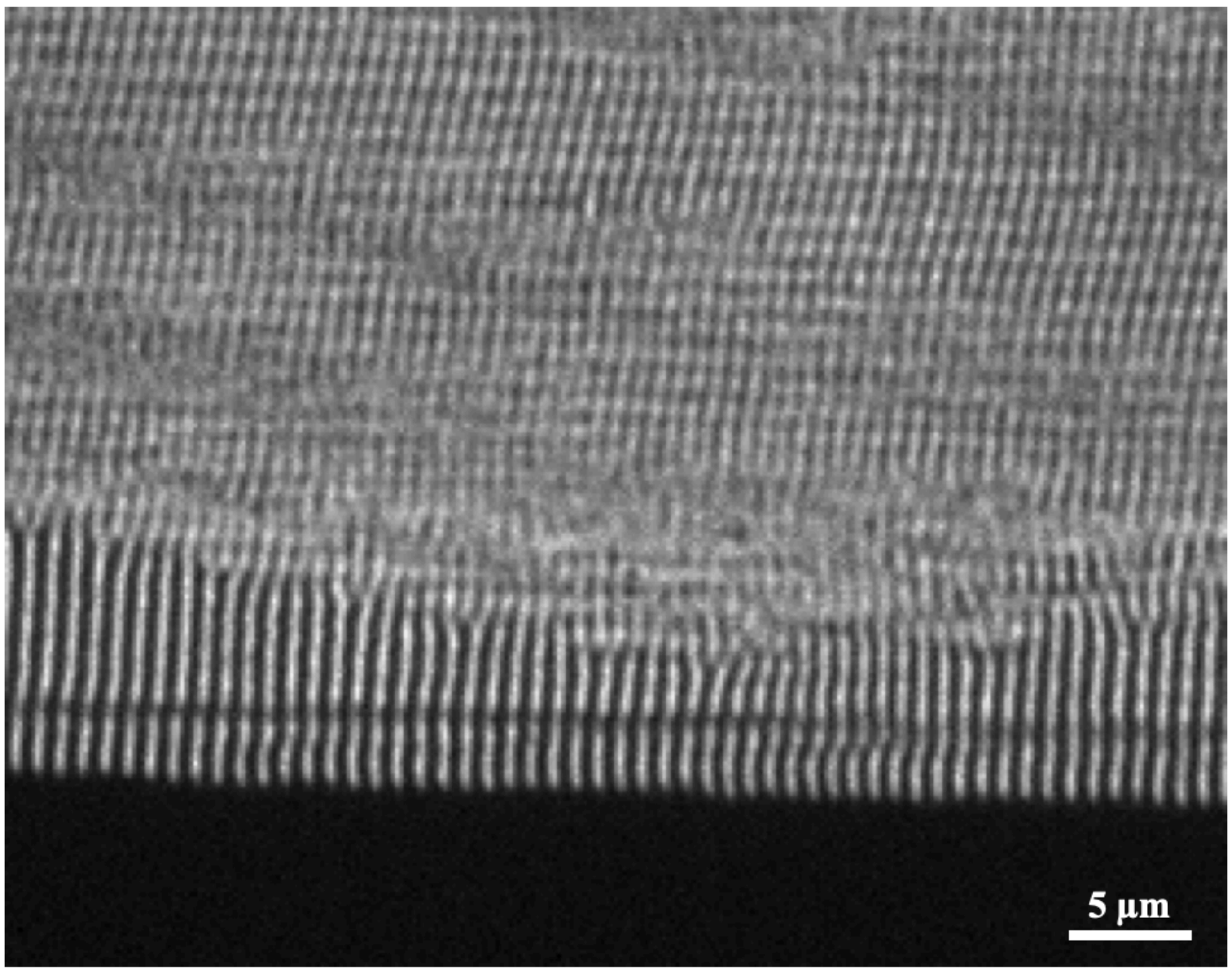

**Figure S5.** Period estimation and measurements of emerging $NLL_{||}$ patterns. Red data points indicate experimentally measured periods, and green points represent values calculated from numerical simulations. Representative samples from both experiment and simulation are displayed below the plot, with green and red frames again distinguishing experimental and simulated patterns, respectively. The simulations accurately predict the variation in the pattern period, demonstrating strong agreement with experimental results. Calculations are performed for titanium. The scale bar represents 5 µm.

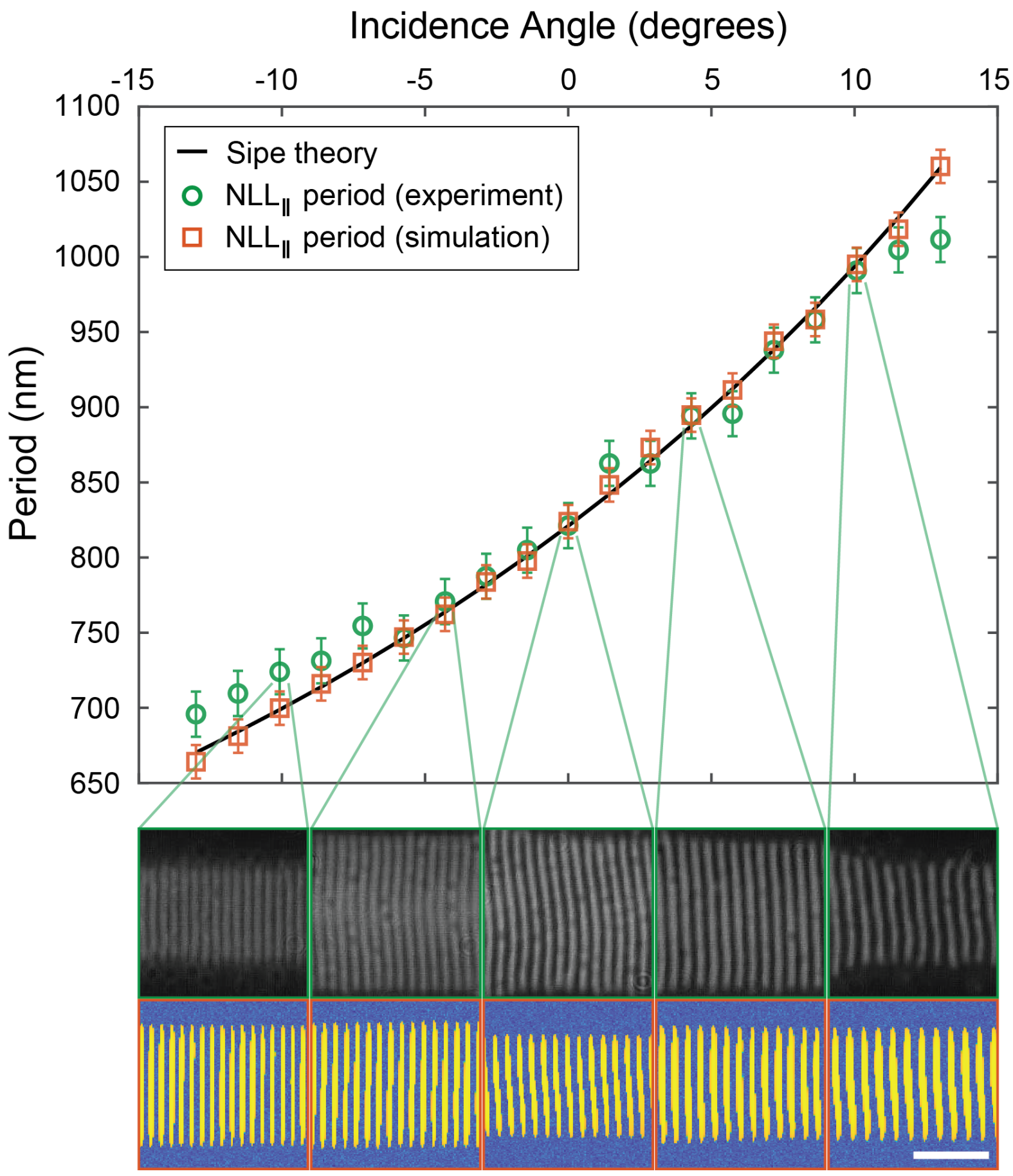

**Figure S6.** Microscopy image showing an example of changing line period during laser scanning.

## 4. Exhaustive demonstration of all primary transitions between the two mechanisms

NLL results in structures in the form of parallel stripes when using linearly polarised laser light. Any area covered by either pattern type has two distinct edges, one where the stripes are parallel and another where they are perpendicular to the edge or boundary. If we allow the use of arbitrarily polarisation angles when creating the two regions, the possibilities increase, in principle, to infinity. We ignore them here and consider only the distinct cases where the laser polarisation is linear and oriented along one direction (say, along the $x$-axis) or orthogonal to it (along the $y$-axis). Even then, there are 12 distinct ways in which $NLL_{\|}$ can transition into $NLL_{\perp}$ or *vice versa*. We have systematically demonstrated all such transitions in Figure S7.

The $NLL_{\|}$ experiments are performed with $N = 2 \times 10^5$, and pulse fluence of 42 mJ/cm². The $NLL_{\perp}$ experiments are performed with $N = 60$, and pulse fluence of 210 mJ/cm². The experiments contain an 8 µm-12 µm overlap region while going from one process to the other only for illustration purposes. While moving from oxidation type to ablation type, the transition region can act as a nucleation site because of the induced irregularities causing some structures to chip away; these irregularities heal further into the ablation region where the chipping behaviour is not observed, and structures are uniform. This effect does not occur from the ablation type to the oxidation type due to the non-destructive naure of the oxidation type NLL process. However, ablated structures can leave debris fields in their vicinity, hampering the coherence of NLL structures newly forming nearby. The oxidation-type NLL structures heal as they move away from the debris field, gaining full coherence. Also, note that the orientation of the ablation-based structures is perpendicular to the laser polarisation.

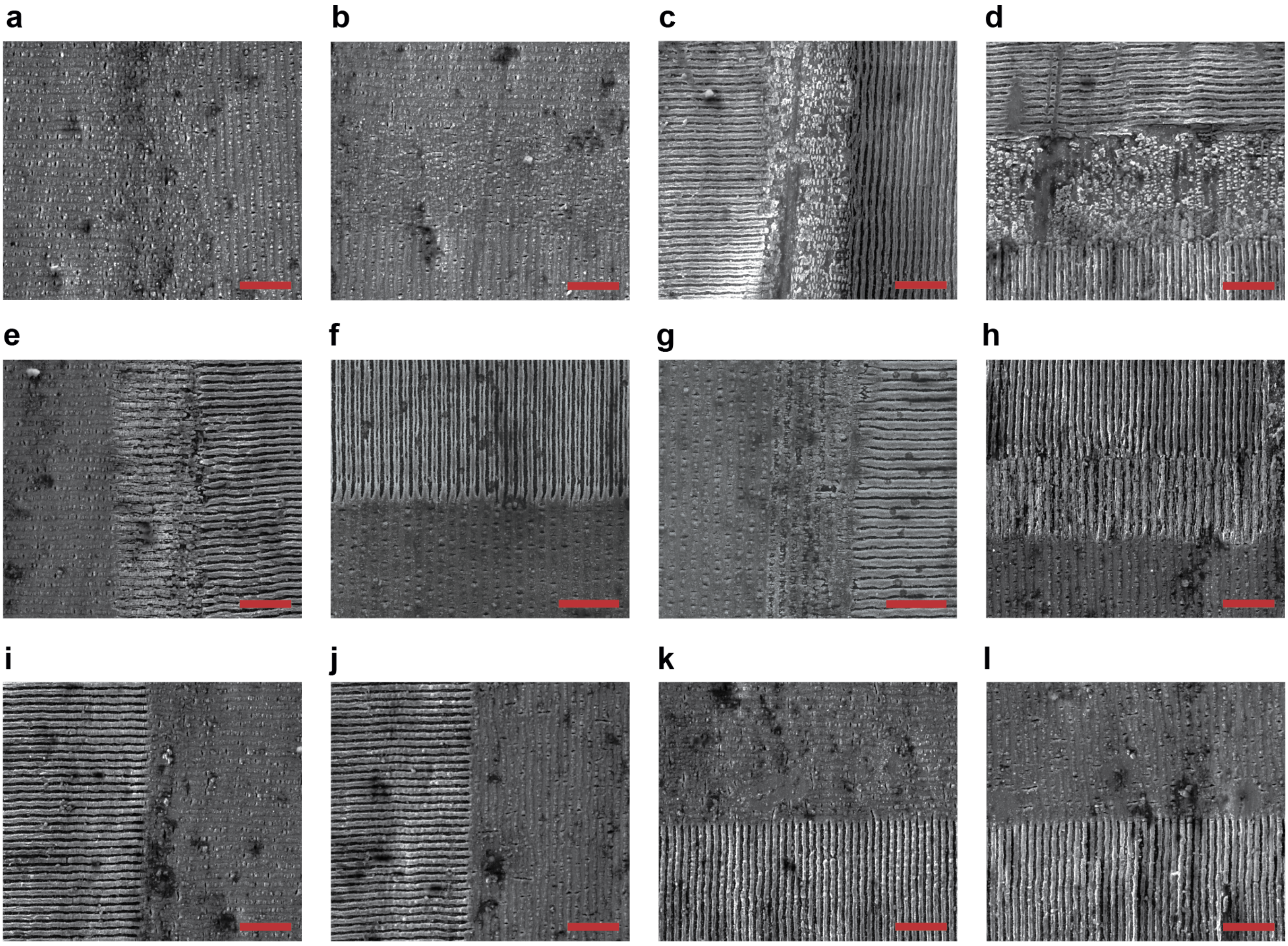

**Figure S7.** The figure shows the possible transitions between $NLL_{||}$ and $NLL_{\perp}$ on cobalt for horizontal and vertical polarisations of the laser beam. **(a)** $NLL_{||}$ horizontal polarisation to vertical polarisation. **(b)** $NLL_{||}$ vertical polarisation to horizontal polarisation. **(c)** $NLL_{\perp}$ vertical polarisation to horizontal polarisation. **(d)** $NLL_{\perp}$ horizontal polarisation to vertical polarisation. **(e)** $NLL_{||}$ horizontal polarisation to $NLL_{\perp}$ vertical polarisation. **(f)** $NLL_{||}$ horizontal polarisation to $NLL_{\perp}$ horizontal polarisation. **(g)** $NLL_{||}$ vertical polarisation to $NLL_{\perp}$ vertical polarisation. **(h)** $NLL_{||}$ vertical polarisation to $NLL_{\perp}$ horizontal polarisation. **(i)** $NLL_{\perp}$ vertical polarisation to $NLL_{||}$ horizontal polarisation. **(j)** $NLL_{\perp}$ vertical polarisation to $NLL_{||}$ vertical polarisation. **(k)** $NLL_{\perp}$ horizontal polarisation to $NLL_{||}$ horizontal polarisation. **(l)** $NLL_{||}$ horizontal polarisation to $NLL_{||}$ vertical polarisation. All scale bars are 5 μm.

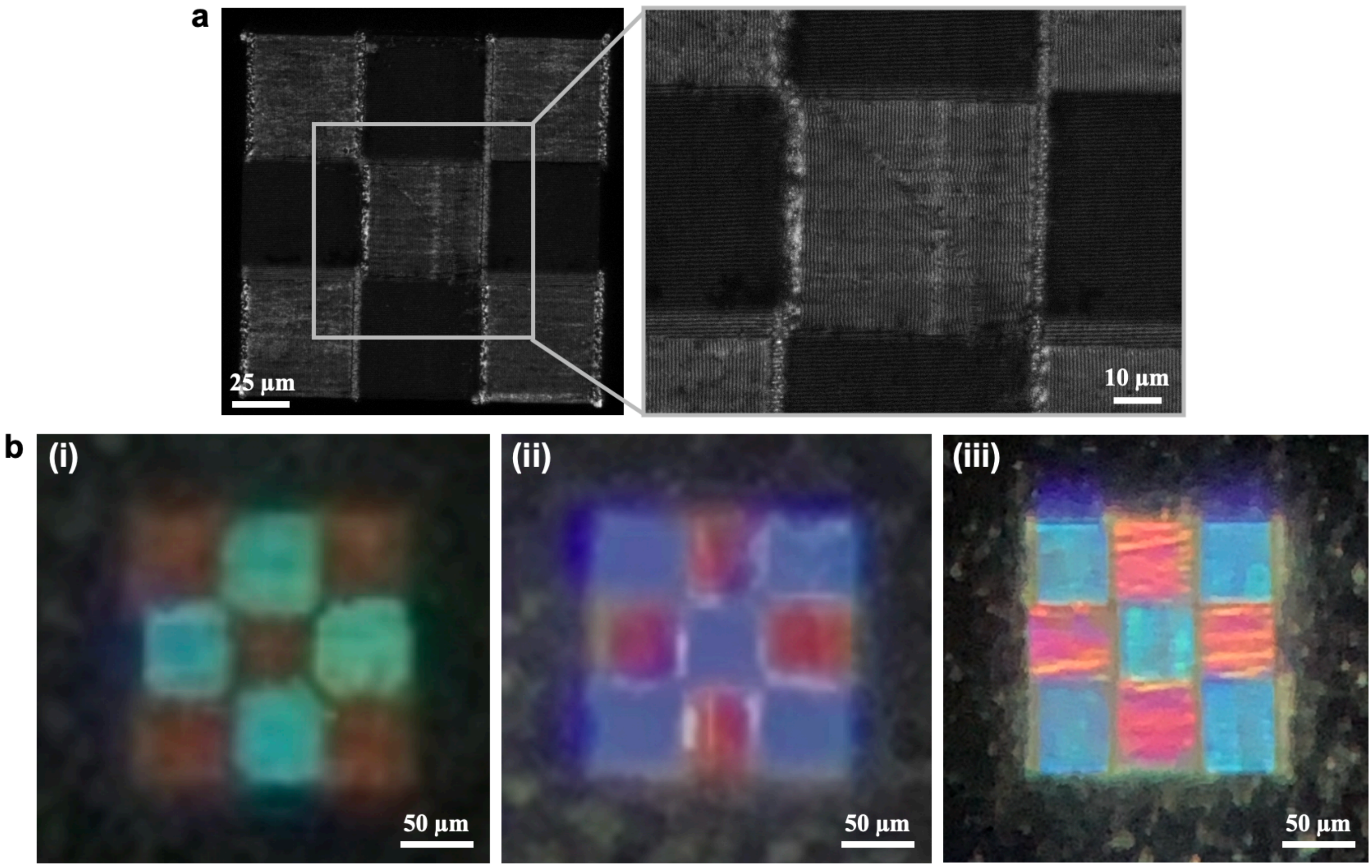

**Figure S8. (a)** Bright-field transmission optical microscope images showing an example of tiling of a surface with alternating NLL patterns on a Co sample. The bright patches are the parts fabricated with $NLL_{\perp}$. Since the coating on the substrate is truly removed from the sample via ablation at the modified points, they look bright. The dark patches are $NLL_{\parallel}$ due to oxide formation above the 100-nm-thick metal coating, where the oxidised parts appear as gray lines. **(b)** Various structural colours appear depending on the illumination angle and rotation of the sample.

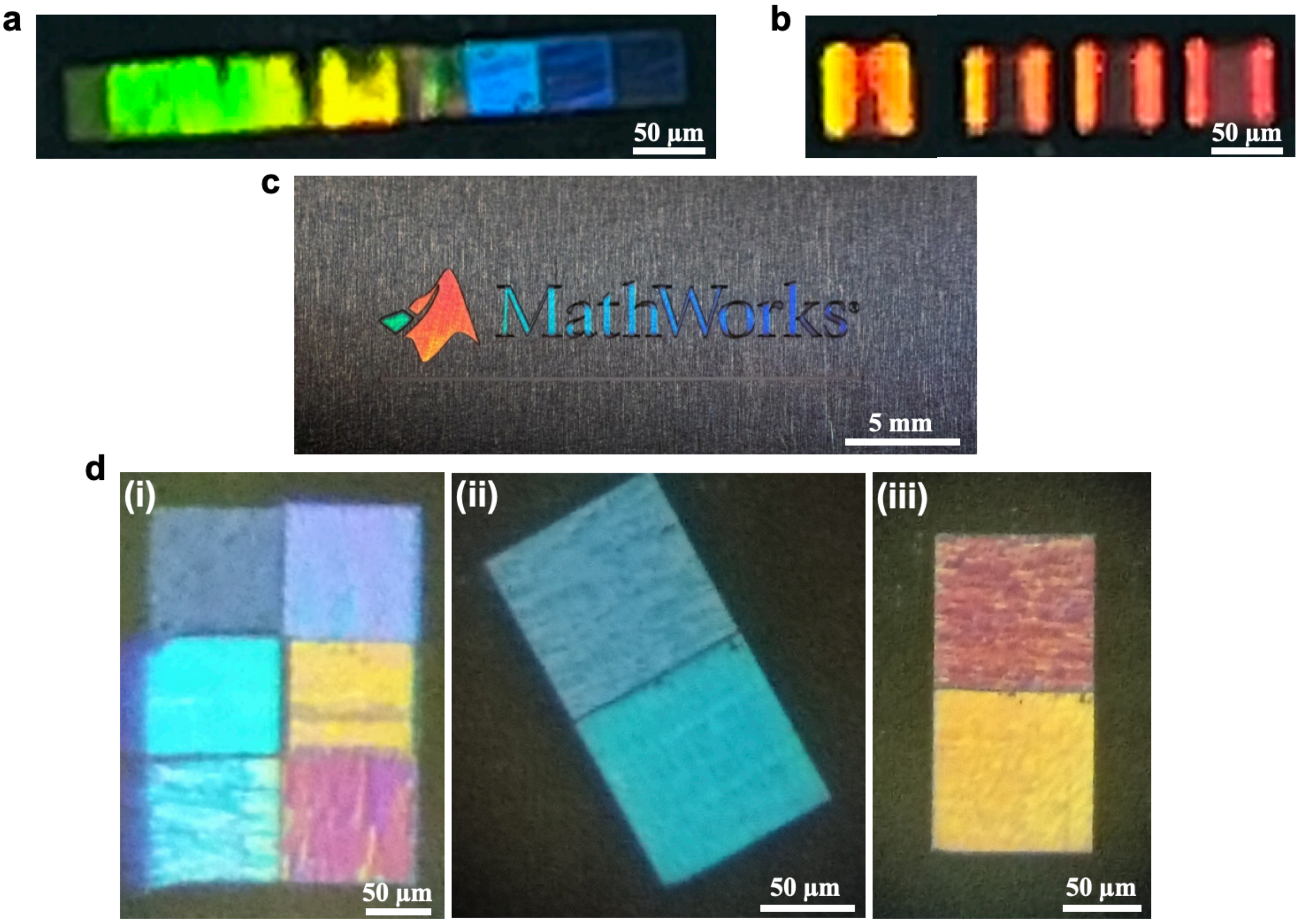

**Figure S9.** Demonstration of the colour palette and its applications. **(a)** $NLL_{\perp}$ with different polarisation on the Co sample. **(b)** $NLL_{\parallel}$ on a Co sample with decreasing applied laser powers. **(c)** Demonstration of controllable colour change on an unpolished stainless steel sample. **(d)** Demonstration of $NLL_{\perp}$ and $NLL_{\parallel}$ illumination angle and rotation of the sample on the Co sample.

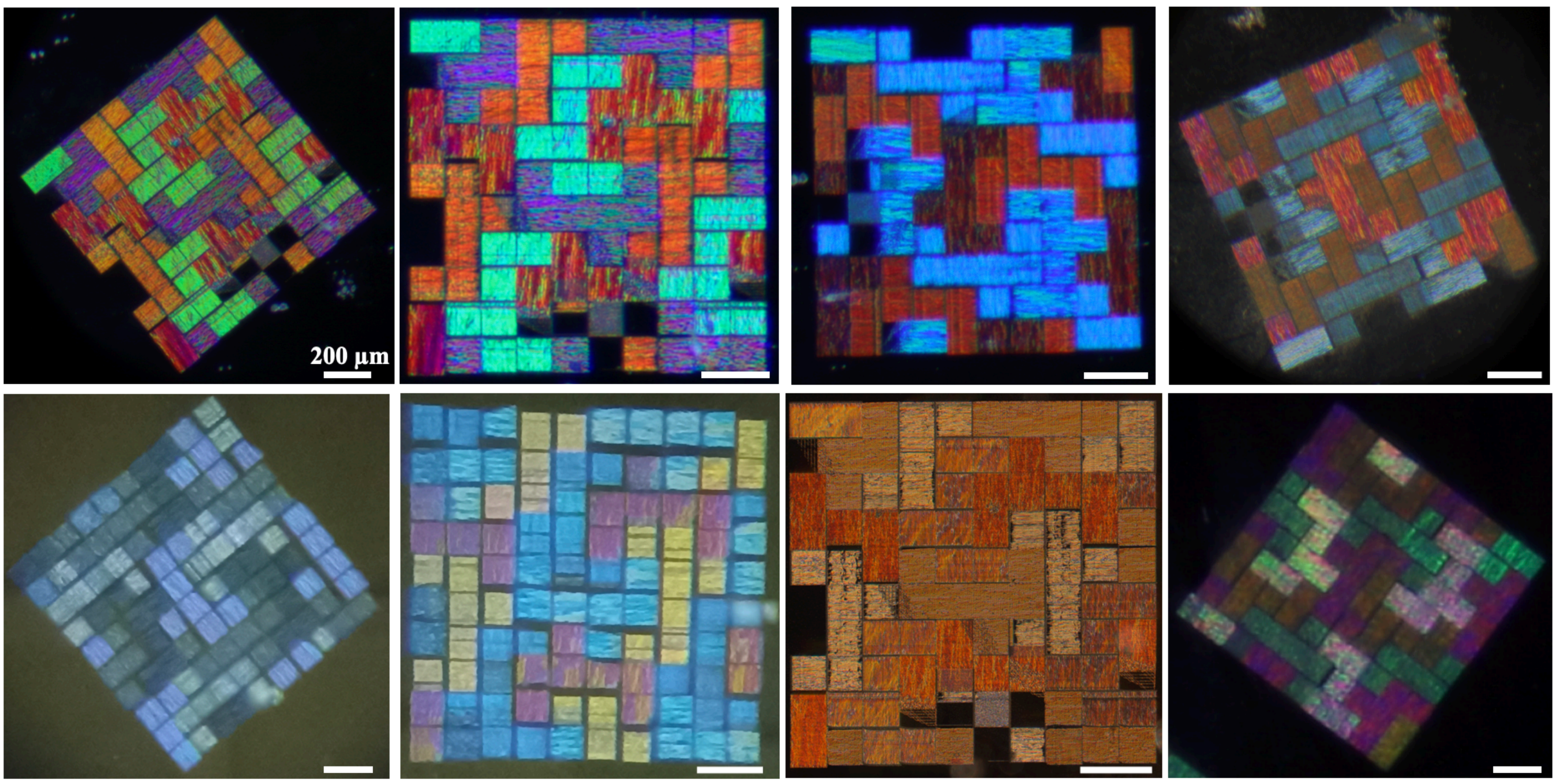

**Figure S10.** Various heptamino tiles with differing colour arrangements, due to the imaging conditions. The orientation of the sample and the angle of white light illumination change the diffraction from each individual cell. All scale bars are 200 µm.

## 5. Demonstration of large-area patterning on flexible glass

In order to demonstrate the large-area processing on flexible materials, we coated a flexible glass (Coresix Willow Glass) with 100-nm-thick titanium. We processed the coated flexible glass using a galvo-scanner. Figure S11 shows the processed area on a bent glass. As the photo shows, the structures can be bent without destroying the pattern. The bending has been repeated many times with no noticeable adverse effects. Even samples, which have been produced several years ago and stored in ambient air at room temperature, maintain the patterned structures and the ability to be bent without any degradation of the structures.

As mentioned in the main text, the processing speed was limited to 7.2 $mm^2/s$ only by the positioning speed of the galvo scanner. Resonant-galvo or polygon scanners allow up to hundreds to a thousand times faster scanning speeds. There is no clear impediment in terms of the NLL process itself to scaling the speeds up to the maximum values allowed by these advanced scanners, and we estimate that 20 $cm^2/s$ should be achievable with an off-the-shelf resonant-galvo scanner.

Therefore, the patterns created by the NLL process are long-durable, do not require special storage conditions, and can be applied at great speeds with strong possibilities for industrial applications.

a b

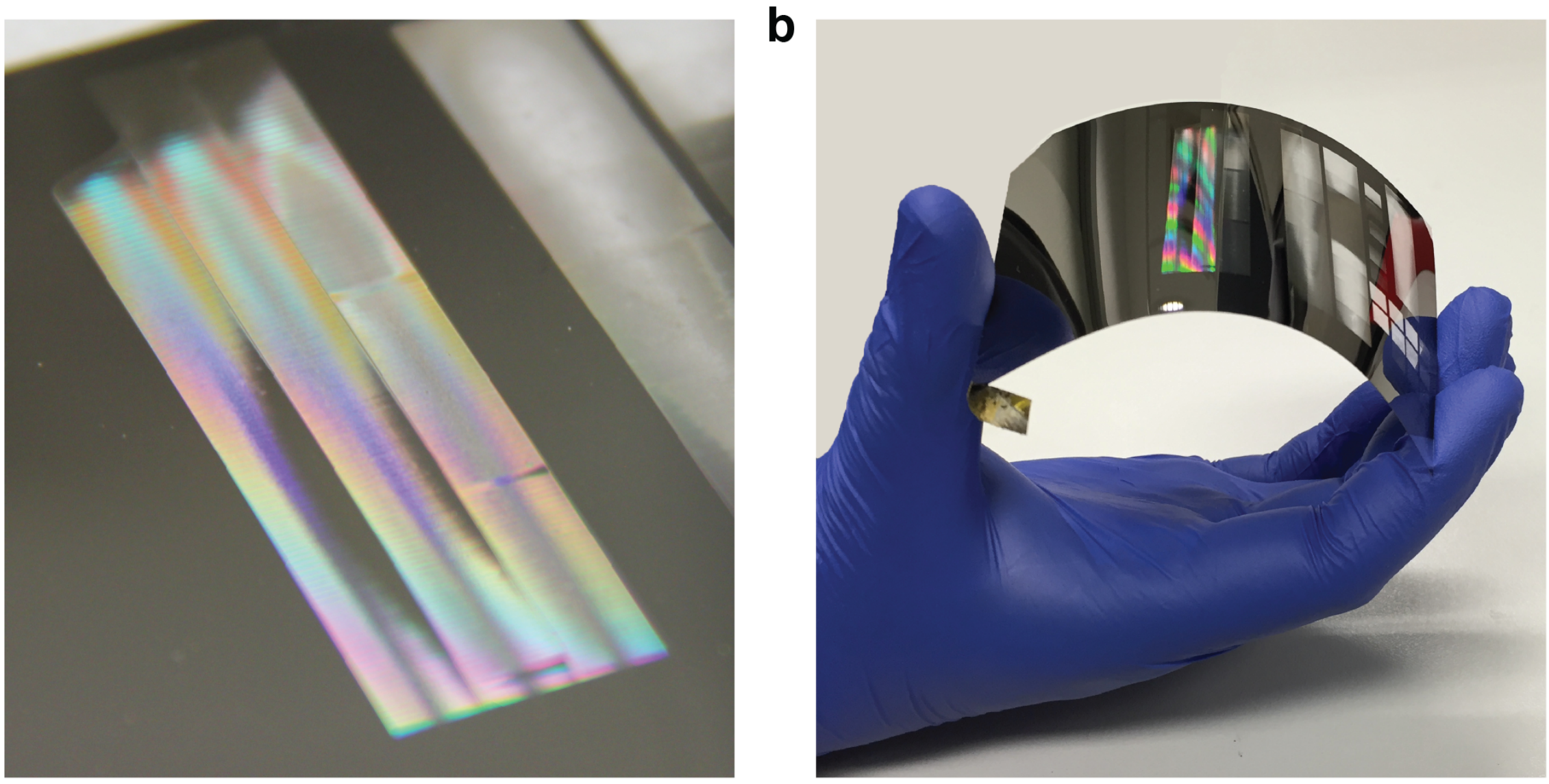

**Figure S11.** Photo of the processed flexible glass. **(a)** Close-up photo of the processed area while the sample is flat. **(b)** Photo of the sample being bent.

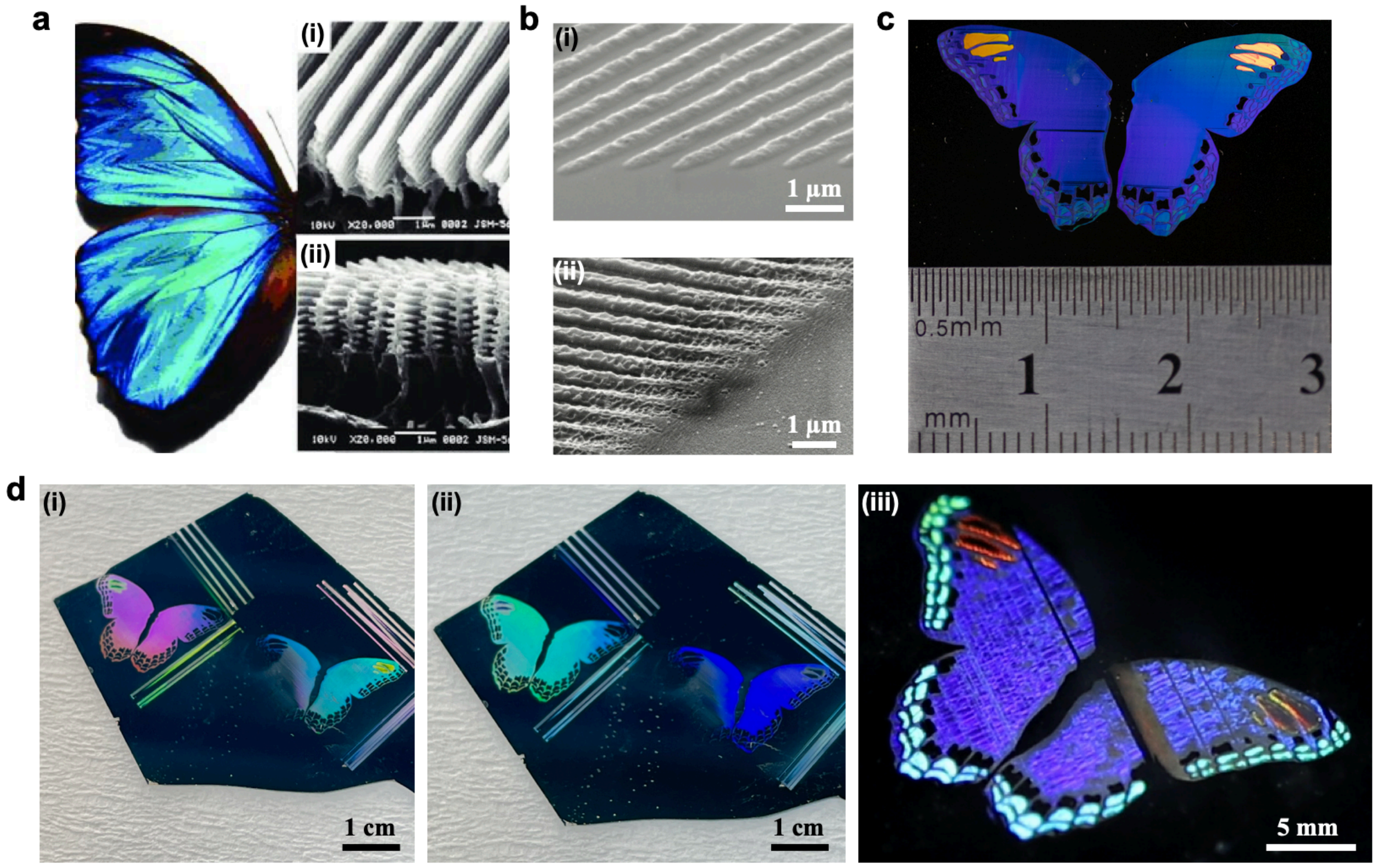

**Figure S12.** Additional images of the butterfly *Limenitis arthemis astyanax* and its depicted by NLL patterns. **(a)** Adapted from [1]. A close-up image of a wing of the real butterfly. SEM images of **(i)** *M. rhetenor* and **(ii)** *M. didius*. **(b)** SEM images of the **(i)** $NLL_{\parallel}$ and **(ii)** $NLL_{\perp}$ patterns. **(c), (d)** Photos of the NLL-created images of the butterfly demonstrate the changing colours with illumination and viewing angle. All structures were created on a 100 nm-thick Co-coated flexible Coresix Willow Glass.